\documentclass{aastex631}

\shorttitle{Grand Design vs.\ Multi-Armed Spiral Galaxies}
\shortauthors{Smith et. al.}
\graphicspath{{./}{figures/}}

\begin{document}

\title{Grand Design vs.\ Multi-Armed Spiral Galaxies: Dependence on Galaxy Structure}

\author[0000-0002-8521-5240]{Beverly J. Smith}
\affiliation{East Tennessee State University \\
Department of Physics and Astronomy, Box 70652 \\
Johnson City TN  37614, USA}

\author{Matthew Watson}
\affiliation{East Tennessee State University \\
Department of Physics and Astronomy, Box 70652 \\
Johnson City TN  37614, USA}

\author[0000-0003-3856-7216]{Mark L. Giroux}
\affiliation{East Tennessee State University \\
Department of Physics and Astronomy, Box 70652 \\
Johnson City TN  37614, USA}

\author[0000-0002-6490-2156]{Curtis Struck}
\affiliation{Iowa State University \\
Department of Physics and Astronomy\\
Ames IA 50011, USA}



\keywords{Disk Galaxies --- Spiral Galaxies --- Galaxy Evolution}


\begin{abstract}

We developed an algorithm to use 
Galaxy Zoo 3D spiral arm masks 
produced by citizen scientist volunteers
to semi-automatically
classify spiral
galaxies as either multi-armed or grand design spirals.
Our final sample consists of 299 multi-armed 
and 245 grand design
galaxies.
On average,
the 
grand design galaxies 
have smaller stellar masses
than the multi-armed
galaxies.
For a given stellar mass, the grand design galaxies have larger 
concentrations, earlier Hubble types,
smaller half-light radii,
and larger central surface mass densities 
than
the 
multi-armed galaxies.
Lower mass galaxies of both arm classes have 
later Hubble types and lower concentrations than higher mass galaxies.
In our sample, a higher fraction of grand design galaxies have classical
bulges rather than pseudo-bulges, compared to multi-armed galaxies.
These results are consistent with theoretical models and simulations
which suggest that dense classical bulges 
support the development and/or longevity of 2-armed spiral
patterns.
Similar 
specific star formation rates are found in
multi-armed and grand design
galaxies with similar stellar masses and concentrations. This implies
that the specific star formation rates 
in spiral galaxies is a function of 
concentration and stellar mass, 
but independent of the number of spiral arms.
Our classifications are consistent with
arm counts from the Galaxy Zoo 2 project and published m=3 Fourier amplitudes.

\end{abstract}

\section{Introduction} \label{sec:intro}

Galaxy morphological classification 
as introduced by \citet{1926ApJ....64..321H} divides spiral galaxies into 
Hubble types based upon the bulge-to-disk ratio,
the tightness of the spiral arms, and the `degree of resolution in
the arms'. 
An alternative method of classifying spiral galaxies  
based upon the number and appearance of the spiral arms
was introduced by the Elmegreens
in a series of papers
\citep{1981ApJS...47..229E, 
1982MNRAS.201.1021E,
1984ApJS...54..127E,
1987ApJ...314....3E,
2011ApJ...737...32E}.
In the modern version of the 
Elmegreen arm class system, grand design
galaxies
have two long continuous spiral arms, flocculent galaxies
have multiple short fragments of spiral arms, and 
galaxies with three or more 
moderately long arms are classified as multi-armed
\citep{2011ApJ...737...32E, 2015ApJS..217...32B}.
Applying both classification schemes in parallel
and examining galactic properties such 
as stellar mass (M*), concentration, bulge structure,
star formation rate (SFR), 
and 
specific SFR (sSFR)
may provide clues to the mechanisms that excite and maintain
spiral arms in galaxies.

Several processes may generate spiral arms in galaxies,
including interactions between galaxies, central bars,
and gravitional instabilities in differentially rotating disks.
In classical spiral density wave theory,
waves with fixed pattern speeds
travel through the disk
\citep{1964ApJ...140..646L, 1966PNAS...55..229L}.
Without some means of rejuvenation, however, spiral 
waves
will disperse within a few galactic revolutions
\citep{1969ApJ...158..899T}.  In theory,
a large bulge may stabilize a spiral density
wave.  
Galaxies with large bulges are expected to have
high Toomre Q parameters in their centers, and a high Q may reflect
incoming waves, producing a 2-armed standing wave that is long-lived  
\citep{1985IAUS..106..513L, 
1989ApJ...338..104B}.
Whether stable wave modes like these occur in real galaxies is uncertain.
\citet{2011MNRAS.410.1637S} found only transient spiral patterns
in a set of simulations based on the 
\citet{1989ApJ...338..104B} scenario, 
but
more recent simulations 
by \citet{2016ApJ...826L..21S}
produced long-lived two-armed
spiral patterns with constant pattern speeds.

Spiral arms may also be caused by 
gravitational instabilities in a
differentially rotating disk 
\citep{1965MNRAS.130..125G, 
1966ApJ...146..810J,
2012ApJ...751...44S, 
2013ApJ...763...46B}.   
In this scenario, the spiral is enhanced by the galactic
shear 
in a process called swing amplification
\citep{1966ApJ...146..810J, 
1981seng.proc..111T, 
2013ApJ...763...46B, 
2013ApJ...766...34D,
2018MNRAS.481..185M,
2014PASA...31...35D}.
Such arms may be self-perpetuating and therefore long-lived
in a statistical sense
\citep{2013ApJ...766...34D}.
Whether the arms are fragmented or continuous may
depend upon the relative amount of stellar feedback;
higher feedback tends to produce a more flocculent structure
in these models
\citep{2018MNRAS.478.3793D}.
Spiral structure may persist longer in disks with on-going
star formation because of dynamical cooling 
\citep{1984ApJ...282...61S}. 

Grand design
spirals may also be driven by galactic bars 
\citep{1976ApJ...209...53S, 1979ApJ...233..539K, 
1980A&A....88..184A}.
Alternatively, 
a gravitational 
perturbation by a neighboring galaxy can induce a 2-armed spiral
pattern
according to models
\citep{1972ApJ...178..623T, 1992AJ....103.1089B,
2010MNRAS.403..625D, 2011MNRAS.414.2498S}.
In addition to M51-like interactions between galaxies
in bound pairs 
\citep{2010MNRAS.403..625D},
rapid flyby encounters may also
induce spiral arms that can be long-lived 
\citep{2008ApJ...683...94O, 2011MNRAS.414.2498S}.

The role of each of these mechanisms in creating the 
observed
spiral patterns of galaxies remains controversial
(see recent reviews by 
\citet{2014PASA...31...35D},
\citet{2016ARA&A..54..667S},
and \citet{2022ARA&A..60...73S}).
To distinguish between these scenarios, 
observational studies in which the number
of arms and/or the arm class is compared with various
properties of the galaxies are helpful.

Past studies show some limited correlation
between
Hubble type and arm class. 
Flocculent galaxies tend to have later Hubble types than 
grand design galaxies, but there is considerable variation
among the Hubble types of galaxies in a given arm class
\citep{1982MNRAS.201.1021E, 
2013JKAS...46..141A,
2017MNRAS.471.1070B, 
2019A&A...631A..94D}.
Galaxies with grand design patterns 
are more likely to have large bulges
\citep{2017MNRAS.471.1070B}.
Flocculent galaxies tend to be smaller than grand design galaxies
\citep{1987ApJ...314....3E} and have lower stellar masses
\citep{2017MNRAS.471.1070B}.

To test spiral arm models observationally, large samples of
galaxies with quantitative estimates of the number of arms are needed.
One way to efficiently extract quantitative measures 
of spiral arm properties for a 
large number of galaxies
is via
Fourier analysis. 
The largest such study done to date is that
of \citet{2020ApJ...900..150Y}, who studied 4378 
face-on (ellipticity $\le$ 0.5) Sloan Digital Sky Survey
(SDSS) disk galaxies with z $\le$ 0.05.  
Using 1-dimensional and 2-dimensional Fourier decomposition
of the SDSS r images, 
\citet{2020ApJ...900..150Y}
derived an average arm strength 
and an average spiral arm pitch angle for each galaxy.
They also calculated
a parameter known as f3, which is the 
radially-averaged 
ratio of the m=3 Fourier 
component amplitude to the sum of the m=2, 
m=3, and m=4 components.   
The f3 parameter is an approximate measure of the number of spiral 
arms
\citep{2011ApJ...737...32E, 2018ApJ...862...13Y}.
\citet{2020ApJ...900..150Y} found that galaxies with larger
central concentrations and earlier Hubble
types tend to have smaller f3 parameters,
implying fewer arms.
In a followup paper comparing the 
\citet{2020ApJ...900..150Y} 
dataset to other parameters,
\citet{2022AJ....164..146S} found that 
f3 is 
weakly positively correlated with sSFR.
However, 
in subsets of the sample
with 
narrow ranges of concentration, this trend disappears.
They conclude that the weak f3-to-sSFR correlation is an 
indirect consequence of
both quantities being inversely correlated with concentration.

Another way to obtain 
estimates of the number of spiral arms in 
large samples of galaxies is via 
the Galaxy Zoo 2
(GZ2) 
citizen scientist survey
\citep{2013MNRAS.435.2835W}.  
In the GZ2 program, participants were asked a 
series of questions
about a set of galaxies, including their best-guess
determinations
of the number of arms in each galaxy.
GZ2 citizen scientist vote fractions 
for the number of arms in galaxies have been used in several 
studies with different galaxy samples, in some
cases with seemingly 
contradictory results.
For a luminosity-limited 
and mass-limited
set of 6683 GZ2 SDSS galaxies, 
\citet{2016MNRAS.461.3663H}
found that the fraction of spiral galaxies that are multi-armed 
increases with stellar mass,
and concluded that multi-armed galaxies have bluer optical colors
than 2-armed galaxies, a difference they attributed to 
enhanced SFRs.
However,
for a different
set of SDSS galaxies, 
\citet{2017MNRAS.468.1850H}
used ultraviolet and mid-infrared 
photometry to measure SFRs 
and 
found little dependence
of SFR and sSFR
on the GZ2 arm count, although possibly higher
dust extinctions in the 2-armed galaxies.
The 
lack of a relation between arm count and sSFR 
seen by \citet{2017MNRAS.468.1850H}
is consistent with earlier results
by \citet{2015MNRAS.449..820W}, who found
that the SFR-M* relation for SDSS galaxies
is independent of GZ2 arm counts.
\citet{2022MNRAS.515.3875P} revisited this topic
in a later Galaxy Zoo 2 project utilizing
more sensitive, better resolution  
Kilo Degree Survey (KiDS) images from the Very Large
Telescope rather than SDSS images.
For a set of 1689 galaxies, 
\citet{2022MNRAS.515.3875P} 
concluded that galaxies
with three, four or five arms tend to have 
higher M* and higher SFRs than galaxies with two
arms, but lower sSFRs. 
In a second GZ2 project using KiDS images, 
\citet{2022MNRAS.517.4575S}
found that the fraction of 
galaxies with
three arms increases as the optical color becomes bluer.

\citet{2023MNRAS.518.1022S}
use a convolutional neural network on SDSS images 
to classify 1354
SDSS galaxies as either grand design or flocculent, 
omitting potential multi-armed galaxies from the sample. 
They trained their code using the \citet{2015ApJS..217...32B}
by-eye classifications of 90 grand design
and 270 flocculent galaxies.
\citet{2023MNRAS.518.1022S} found that the flocculent
galaxies identified by their neural network code
have lower rotational velocities (lower dynamical masses)
and later Hubble types than the galaxies classified as grand
design.

As discussed above, the relation of the number of arms
to other basic galaxian quantities 
such as M*, SFR, sSFR, and concentration
remains unsettled.
This uncertainty impacts the question of how spiral arms form
and evolve with time.
Since models of spiral production via disk instabilities and
swing amplification
tend to produce galaxies with flocculent or multi-armed
structure
\citep{2013ApJ...763...46B, 
2013ApJ...766...34D,
2018MNRAS.478.3793D}, 
it is sometimes assumed
that grand design galaxies
are produced by interactions or bars, while 
gravitational instabilities produce
other spirals
\citep{2014PASA...31...35D}.
However, instability models can 
also 
produce 2-armed 
morphologies
if the disk/halo mass ratio is high enough
\citep{1984ApJ...282...61S,
2015ApJ...808L...8D,
2018MNRAS.481..185M}.
According to both swing amplification theory and numerical
simulations,
galaxies with disks that are more massive 
relative to the halo mass tend to 
have fewer arms
\citep{1985ApJ...298..486C, 2013ApJ...766...34D, 
2015ApJ...808L...8D,
2018MNRAS.478..932H, 2018MNRAS.481..185M}.
The number of arms may also depend upon whether the
distribution of 
dark matter in the halo is `cuspy' or `flat'
\citep{2018MNRAS.478..932H}
and on the galactic shear rate
\citep{2018MNRAS.481..185M}.

Supporting the theory that grand design spirals are produced by 
interactions,
observational studies have found higher proportions
of grand design galaxies among binary galaxies and group galaxies
compared to field galaxies
\citep{1982MNRAS.201.1021E}.  Furthermore,
clusters have higher
proportions of grand design galaxies compared to the field
\citep{1982MNRAS.201.1035E,
2011JKAS...44..161C}.
Using GZ2, 
\citet{2016MNRAS.461.3663H}
found increasing fractions of 2-armed galaxies in higher
density regions.  
\citet{2020MNRAS.493..390S}
also concluded that
non-isolated galaxies are more likely to be grand design compared
to isolated galaxies.
In agreement with these studies,
\citet{2022AJ....164..146S}
found lower normalized m=3 Fourier components 
on average for cluster galaxies than for galaxies in the field,
implying a higher percentage of 2-armed galaxies in clusters.
However, they attributed
their result to galaxies in clusters having larger concentrations
than field galaxies.   
When they compared
galaxies with the same concentration and the same mass,
the distribution of f3 values are the same in clusters as in the field.
They thus concluded that the number of arms in a galaxy
is a product of the internal structure of the galaxy and not the environment.

Observational tests of bar-driven spirals
are inconclusive so far.
Bar strength and arm strength tend to be correlated
\citep{2004AJ....128..183B, 2005AJ....130..506B, 2020ApJ...900..150Y},
however, this correlation may mean that conditions 
in a disk that favor bar production also favor strong arms,
rather than bars causing the arms
\citep{2010ApJ...715L..56S, 2019A&A...631A..94D}.
In any case, the presence of grand design spiral arms in galaxies
that are both isolated and unbarred highlights the need for
other means of spiral arm production. 

In contrast with the suggestion by 
\citet{2014PASA...31...35D}
that grand design patterns are produced by interactions or bars,
\citet{2017MNRAS.471.1070B} 
and 
\citet{2020ApJ...900..150Y} 
cite the theoretical work of
\citet{1989ApJ...338..104B} 
on the reflection of waves by large 
bulges,
and conclude 
that 
long-lasting wave modes are responsible
for the spiral patterns
in high concentration grand design galaxies.
This disagreement in the literature illustrates 
the lack of consensus on the primary mechanisms
triggering and maintaining spiral arms in galaxies.

In the current paper, 
we 
investigate 
these questions by deriving
arm
counts for a large sample of galaxies
using a different method.
We use spiral arm
masks from a different Galaxy Zoo citizen science project,
the Galaxy Zoo 3D (GZ-3D) project\footnote{https://www.zooniverse.org/projects/klmasters/galaxy-zoo-3d}
\citep{2021MNRAS.507.3923M},
to count the number of arms 
in a sample of SDSS spiral galaxies.  
In Section \ref{sec:sample}, we describe the
sample of galaxies, explain the 
GZ-3D spiral arm masks,
and describe the algorithm we use
to count the spiral arms.  We compare our arm counts
with other properties of the galaxies (stellar mass, concentration,
Hubble type, central surface mass density,
and
sSFR) in Section \ref{sec:results}.  
In Section \ref{sec:published}, we compare our arm counts with 
the \citet{2020ApJ...900..150Y}
arm parameters and
arm counts from the Galaxy Zoo 2 project.
The results are discussed in terms of models of spiral
arm production 
and galaxy evolution in Section \ref{sec:discussion}.
Conclusions are given in Section \ref{sec:summary}.
In 
Appendix \ref{sec:AppendixA},
we 
show GZ-3D masks and SDSS images
of example 
grand design and 
multi-armed galaxies in our sample.
In Appendix \ref{sec:selection}, we discuss possible selection effects
and investigate whether these affect our final conclusions.

Throughout this paper, we assume a Hubble constant of
70 km~s$^{-1}$~Mpc$^{-1}$.

\section{Sample and Data} \label{sec:sample}

\subsection{Sample} \label{sec:sample_subsection}

The 
GZ-3D sample 
\citep{2021MNRAS.507.3923M}
consists of 29,831 
SDSS galaxies with z $<$ 0.15.  The GZ-3D galaxies are a subset of the 
target list for the SDSS-IV
Mapping Nearby Galaxies at APO
(MaNGA) integral field unit (IFU) survey, which in turn was selected
from the NASA Sloan Atlas (NSA) Version 1.0.1 
\citep{2017AJ....154...86W}.
The MaNGA target sample (and therefore the GZ-3D sample) is not
a complete sample, but instead was selected to produce approximately
uniform coverage in log M*, and have suitable angular sizes 
for the MaNGA IFU field of view.
The `Primary' MaNGA sample (about 50\% of the MaNGA sample) 
includes galaxies covered by the 
MaNGA IFU field out to 1.5 $\times$ the effective radius of the galaxy,
while the `Secondary' MaNGA sample are galaxies observed by the MaNGA IFU
out to 2.5 effective radii.  
A third `Color-Enhanced' sample, making up 17$\%$ of the targets,
includes galaxies under-represented in the NUV $-$ i 
color vs.\ 
absolute i magnitude diagram, such as high mass blue
galaxies and low mass red galaxies.

We limit our sample to galaxies with 
z $<$ 0.05,  to allow for 
sufficient resolution 
to accurately measure the spiral arms.
This is approximately the limit at which one can distinguish
grand design from multi-armed galaxies in SDSS images, 
according to simulations of SDSS images by 
\citet{2018ApJ...862...13Y}.
We also limit our sample to 
relatively face-on galaxies with 
ellipticity in the NSA $\le$ 0.5,
where
ellipticity = 1 $-$ b/a, and b and a are the 
semi-minor and semi-major axis of the galaxy, respectively.
This gives us an initial sample of 
16,655 galaxies
with redshifts between 0.01 and 0.05, 
and a median redshift of 0.030.
This corresponds to distances between 43 and 214 
Mpc.

We cross-correlated the \citet{2020ApJ...900..150Y} galaxy list
with the 
\citet{2021MNRAS.507.3923M}
Galaxy Zoo-3D sample.
A total of 1002 galaxies are in both the 
GZ-3D and \citet{2020ApJ...900..150Y}
samples.
In
Section
\ref{sec:YuHo} of this paper, 
we compare our arm 
parameters with 
those of \citet{2020ApJ...900..150Y}.

\subsection{GZ-3D Masks} \label{sec:datamasks}

In the GZ-3D citizen science project 
\citep{2021MNRAS.507.3923M},
participants marked spiral arms and bars on the 
SDSS images of the GZ-3D galaxies.  
GZ-3D participants
also marked the galaxy centers and foreground stars.  
For the spiral arm marking activity in
GZ-3D, the targets were chosen 
from the subset of the MaNGA target galaxies previously 
classified in the earlier Galaxy Zoo 2 project to be spiral 
galaxies with $\le$ 4 spiral arms.  This selection
criteria eliminates some of the most flocculent spiral galaxies
as well as most ellipticals.  This limits the sample
of galaxies with available GZ-3D spiral arm masks
to only 7418 out of the 29,831 galaxies in the full GZ-3D sample
\citep{2021MNRAS.507.3923M}.
Of these 7418 galaxies, 4449 meet our criteria
of z $<$ 0.05 and ellipticity $\le$ 0.5. 
This set of 4449 galaxies is called our low z GZ-3D spiral arm
sample in the following.

Each galaxy in this list 
was viewed by at least 15 GZ-3D volunteers.  
Spiral arm masks were made
for each galaxy based on the markings by the viewers.  
In the mask, each pixel in the image is set to an
integer number between 0 to 15, where the count is equal to the 
number of volunteers that identified that pixel as belonging to
a spiral arm.  
Similar masks were made for the volunteer markings of the bars.
The GZ-3D masks have pixels 0\farcs099 wide,
and are 525 $\times$ 525 pixels (52$''$ $\times$ 52$''$).  
This corresponds to a field of view of 11 kpc (54 kpc) on a side
for the closest (most distant) galaxies.

\subsection{Algorithm for Finding and Counting Arms} \label{sec:algorithm}

We use the GZ-3D masks to identify spiral arms in the galaxies.
In our analysis, for a given pixel to count as being 
`on the arm', we use a threshold of at least three counts, 
i.e., out of the 15 possible viewers, at least three participants 
(20\%)
selected that pixel as being on an arm.
This threshold is recommended by 
\citet{2021MNRAS.507.3923M},
who noted that this cutoff reveals continuous and distinct spiral
patterns.  We consider pixels in the `interarm' 
as pixels which none of the participants labeled as being 
on the arm (i.e., count = 0).
This is a conservative limit.  The consequences of these
criteria are discussed below.

We converted the pixel x,y coordinates into polar coordinates 
centered on the galaxy center, using the central position
marked by the GZ-3D participants.  
When two or more galaxies are in the field of view 
and the nuclei were both marked by volunteers,
we used the nucleus closest to the center of the image.
We then stepped out in 
radius from that center in elliptical annuli,
deprojecting the mask into the plane of the galaxy.
We assume that the intrinsic shape
of the disk in the plane of the galaxy is circular,
and these elliptical annuli are projections from 
the plane of the sky to circles in the plane of the galaxy disk.  
For this deprojection, 
for the galaxies in the \citet{2020ApJ...900..150Y} sample,
we used their determination of ellipticity 
and position angle rather than the NSA value as they are more 
accurate
\citep{2020ApJ...900..150Y}.
If the galaxy was not in \citet{2020ApJ...900..150Y},
then we used the NSA elliptical Petrosian PA and 
ellipticity.  
For each x,y
position on the image, 
we calculated the azimuthal angle $\theta$ in the plane of the 
sky and the distance from 
the galaxy center in the plane of the galaxy.

We then stepped out in 
radius from the center in one pixel wide radial bins, 
searching for arms along each annulus.
For each radial step we searched in azimuthal angle along 
the annulus, looking for pixels that exceed our nominal 
threshold level.  After identifying an arm, we continue along 
the annulus, searching for a pixel that meets our interarm criteria.   
After finding interarm, we continue along the annulus 
searching for another pixel that exceeds the threshold for 
an arm, followed by an interarm pixel.  Each independent 
`arm/interarm' sequence along the annulus is defined as an 
independent arm at that radius.  This means that each arm 
must have defined interarm pixels on either side of the arm 
in the annulus.  If the entire annulus is above the interarm 
level of zero counts, then the arm count for that annulus 
is zero.
Alternatively, if no pixels along
the annulus exceeds the threshold, the arm count for that 
annulus is also zero.   To identify an arm, there must be
at least one pixel along the annulus at or above the threshold, 
and at least one pixel at the defined interarm level.

In stepping out in the galaxy, we started at a radius of 
20 pixels (1\farcs98) from the center, and continued to a radius 
of 250 pixels (24\farcs75), stepping out 1 pixel at a time.
In our final analysis, we only flag galaxies 
which have at least 10 radial steps 
with at least one arm by
the above criteria.   This means that galaxies with weak or 
no spiral arms, or very flocculent or ill-defined arms, 
are unflagged.  
Among the galaxies with ten radial steps with arms, we 
flagged a galaxy as a 
possible n-armed galaxy if at least four of the radial 
steps in a continuous sequence (i.e., at least 0\farcs396 in radius)
were found to have n arms by the above criteria, where n = 1,2,3,4,
and $\ge$4.
Note that the same galaxy can be flagged in more than one
way
if we find different arm counts in different 4-step sequences.
An example would be a 
galaxy that is flagged as 2-armed for a given range of radius,
and 3-armed for a different range of radius.
Out of our initial sample of 4449 low z GZ-3D spiral arm galaxies, the number
of galaxies flagged as 1-armed, 2-armed, 3-armed, 4-armed, and 5+arms,
are 3965, 2806, 367, 23, and 3, respectively.

We flagged a subset of the `2-armed' 
galaxies as `grand design'
spirals, if: a) only two arms were found in a 
continuous sequence of at least 50 
radial steps (4\farcs95),
and
b) the galaxy was not flagged as 3-armed or 4-armed.
Based on these criteria, 
we flagged a total of 
402 galaxies as grand design.

In this section, we have outlined our attempt to use the superior
pattern finding abilities of human observers to develop a consensus
classification scheme.  In the next section, we assess the reliability
of our classifications.

\subsection{Visual Inspection of Images} \label{sec:inspection}

We used the ds9 software\footnote{https://sites.google.com/cfa.harvard.edu/saoimageds9}
to inspect by eye the SDSS g-band images
of all of the galaxies flagged 
by our code as either grand design or as  
3-armed.  Since spiral arms show up more readily
in blue images than red images, we used the g images
for these inspections.
Using ds9, we varied the greyscale display by hand
for each galaxy, and inspected each galaxy carefully.
We found that in some cases
the GZ-3D spiral arm masks at a threshold = 3 
do not trace arms as far out as can been seen by
eye in the 
SDSS g image.  Furthermore, sometimes multiple 
arms can be
seen by eye in the galaxy, but not all of these
arms are above the threshold=3 level in the GZ-3D mask.
Also, in some cases a single GZ-3D volunteer marked a large
portion of the disk as `arm', causing arms in that
region not to be found by our algorithm due to the
lack of `interarm' pixels in the mask.

However, from our inspection of the images 
we conclude that the arms that are found from
the GZ-3D masks using our method with  
a threshold of 3 are generally reliable.
The main shortcoming of our method is that sometimes
additional fainter arms are missed, 
or the arms aren't traced
out as far on the image as can been seen by eye.
That means that the arm counts from our software
may be lower limits on the true arm counts for some galaxies.
Inspection shows that most galaxies flagged 
as 3-armed by our code have 
at least three arms, and sometimes more arms.
In other words, they comprise a robust sample of multi-armed
galaxies.

We explored how our results would change if we set our definition
of interarm to 1 count instead of zero.  With this more relaxed
definition, the code is able to recover more spiral arms,
however, it sometimes gives
false positives, i.e., it sometimes counts more 
arms than are visible by eye.  Thus, for our final analysis
we used the interarm = 0 criterion, which is more reliable
although less complete.

The images were independently inspected by three of the co-authors.
We
divided them by eye into three groups:
galaxies that 
`very likely' belonged to the arm class the code selected, 
galaxies that `likely'
belonged to the class, and galaxies that may not belong to the class.
The three sets of by-eye classifications were then merged into a single
list.  If two or all three of the by-eye rankings of a particular galaxy
agreed, then the galaxy was assigned to that group.  
If the three by-eye rankings were different,
then the galaxy was placed in the middle `likely' group. 
The individual by-eye rankings were reasonably consistent, 
with 
92\% of the galaxies having at least two of the three rankings
the same.
In comparing the properties of the multi-armed and grand design
galaxies, 
we combine the `very likely' and `likely' galaxies together
to make the final samples of galaxies.
The final conclusions of this paper do not change if only
the `very likely' galaxies were included in the analysis.

Of the 367 galaxies flagged by the code as 3-armed,
by eye
we classified 202 (55\%) as `very likely' multi-armed
and another 97 (26\%) as `likely' multi-armed.
Together, this 
gives an overall estimated reliability of
81\%
for our automatic 
method of finding multi-armed galaxies, 
assuming that the consensus vote of our by-eye classifications is
the correct classification.   In other words, 
the code selects a sample of 3+ armed
galaxies that is 85\% `pure' multi-armed galaxies.
The remaining 15\% are mostly galaxies that are 
dominated by 2-arms that are erroneously classified.
These misclassifications include 
barred galaxies
with two tightly-wrapped arms in a ring-like structure, 
where the code erroneously
counted one arm twice, perhaps because the
NSA 
ellipticity is somewhat inaccurate or the arm is distorted.
After filtering out these incorrectly-classified galaxies by eye,
our final set 
of visually-confirmed multi-armed 
galaxies contains 299 galaxies.  We use these 299 galaxies
in the following analysis.

Our sample of multi-armed galaxies includes galaxies with 
branching spiral arms, for example, galaxies with
two dominant arms in the interior that bifurcate to produce
four long arms in the outer disk.  This morphology is common among
galaxies classified as multi-armed
\citep{2015ApJS..217...32B}.  The Milky Way may have
this structure
\citep{2021A&A...654A.138M, 2023ApJ...947...54X}.

Because the GZ-3D masks sometimes miss spiral arms,
the code sometimes under-counts the number of arms.
This means that 
galaxies that are only flagged as 
`2-armed' might in fact have three or more arms.
Our strict definition of grand design galaxies, in
which we require
50 radial steps in sequence
with only 2 arms and exclude galaxies also flagged as 3-armed, 
improves the reliability
of the classification, but it is not perfect.
Of the 402 galaxies selected as grand design by
our algorithm, 146 (36\%) were seen by eye as
`very likely' grand design, and 99 (24\%) were
classified as `likely' grand design, implying 
a 60\% overall reliability.
Some of these galaxies show 
short branching of the spiral
arms in the outer disk, but overall they are dominated by
two arms over the majority of the disk, thus we
classify them as grand design for this study.
The final set of 245 galaxies confirmed by eye to
be likely/very likely grand design 
is used in the analysis below.

Although our method of finding multi-armed and grand design galaxies
using the GZ-3D masks
is reasonably reliable, the final samples are very incomplete.
Out of the 4449 galaxies in the z $<$ 0.05, ellipticity $<$ 0.5 
subset of the GZ-3D spiral arm sample,
3994 (90\%) were initially flagged by our procedure.
Of those 3994 galaxies, 
only a fraction (14\%) ended up in 
either 
our final sample of 299 multi-armed galaxies or the final sample of 
245 grand
design galaxies. 
This incompleteness should be kept in mind in the following
analysis.

In Appendix \ref{sec:AppendixA} of this paper, we provide some examples
of galaxies identified as grand design and as multi-armed, 
show
the GZ-3D masks for these galaxies,
and 
illustrate our flagging procedure.
In Appendix \ref{sec:selection}, we investigate whether
the limited field of view biases the final conclusions of this paper.

\section{Results} \label{sec:results}

\subsection{Arm Counts vs.\ Redshift and M*} \label{sec:stellarmass}

\begin{figure}[ht!]
\plotone{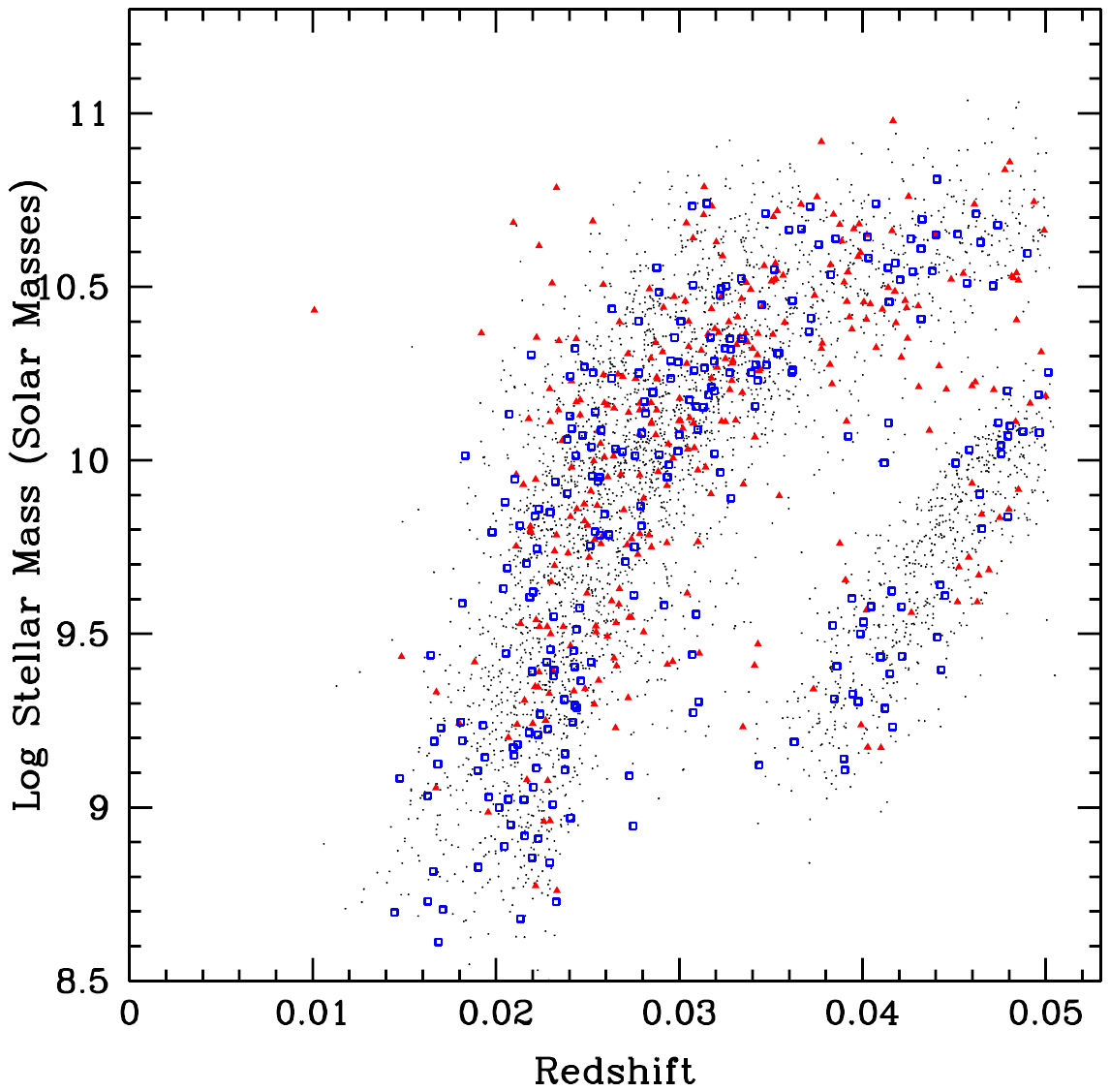}
\caption{ 
Stellar mass vs.\ redshift for our
grand design (blue open squares)
and 
multi-armed galaxies (red filled triangles).  Other 
galaxies in our initial set of 4449 GZ-3D galaxies
with 
z $<$ 0.05, ellipticity $\le$ 0.5 
are marked by small black dots.
The galaxies in the `Primary' and `Secondary' MaNGA subsets
are separated into two `bands' in this plot, with the larger
angular size `Primary' galaxies in the leftmost band. 
\label{fig:redshift_vs_mass}}
\end{figure}

The MaNGA/GZ-3D selection criteria is illustrated 
in Figure \ref{fig:redshift_vs_mass},
where we plot stellar mass (obtained from the NSA) vs.\ redshift.  
The galaxies in our final grand design sample are
shown as
blue open squares, and the multi-armed galaxies
are marked 
by red filled triangles.  The small black dots represent
galaxies in the low z GZ-3D spiral arm sample that are not 
in our final GD or MA sample.
Galaxies in the larger
angular size `Primary' MaNGA subset
populate a broad band of galaxies extending
from redshifts of about 0.02 at low masses 
to a redshift of 0.05
at high masses.   The second band of galaxies at higher
redshifts includes galaxies in the `Secondary' sample,
which have smaller angular sizes. 
The MaNGA `Color-Enhanced' add-on sample of galaxies
contributes to the
scatter in the plot.   
Both grand design and multi-armed galaxies are seen in the
full range of masses
and redshifts of the sample.  

\begin{figure}[ht!]
\plottwo{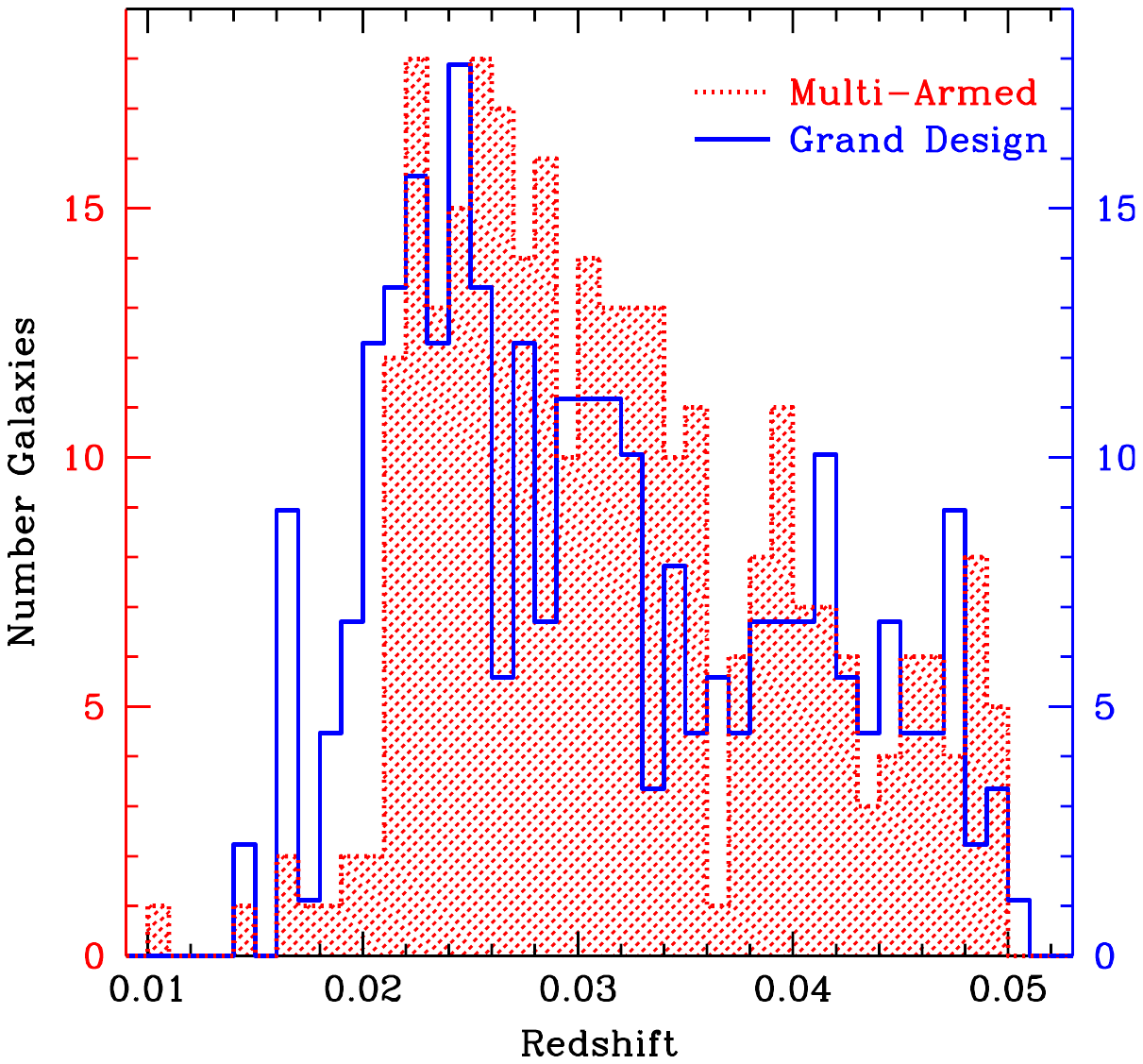}{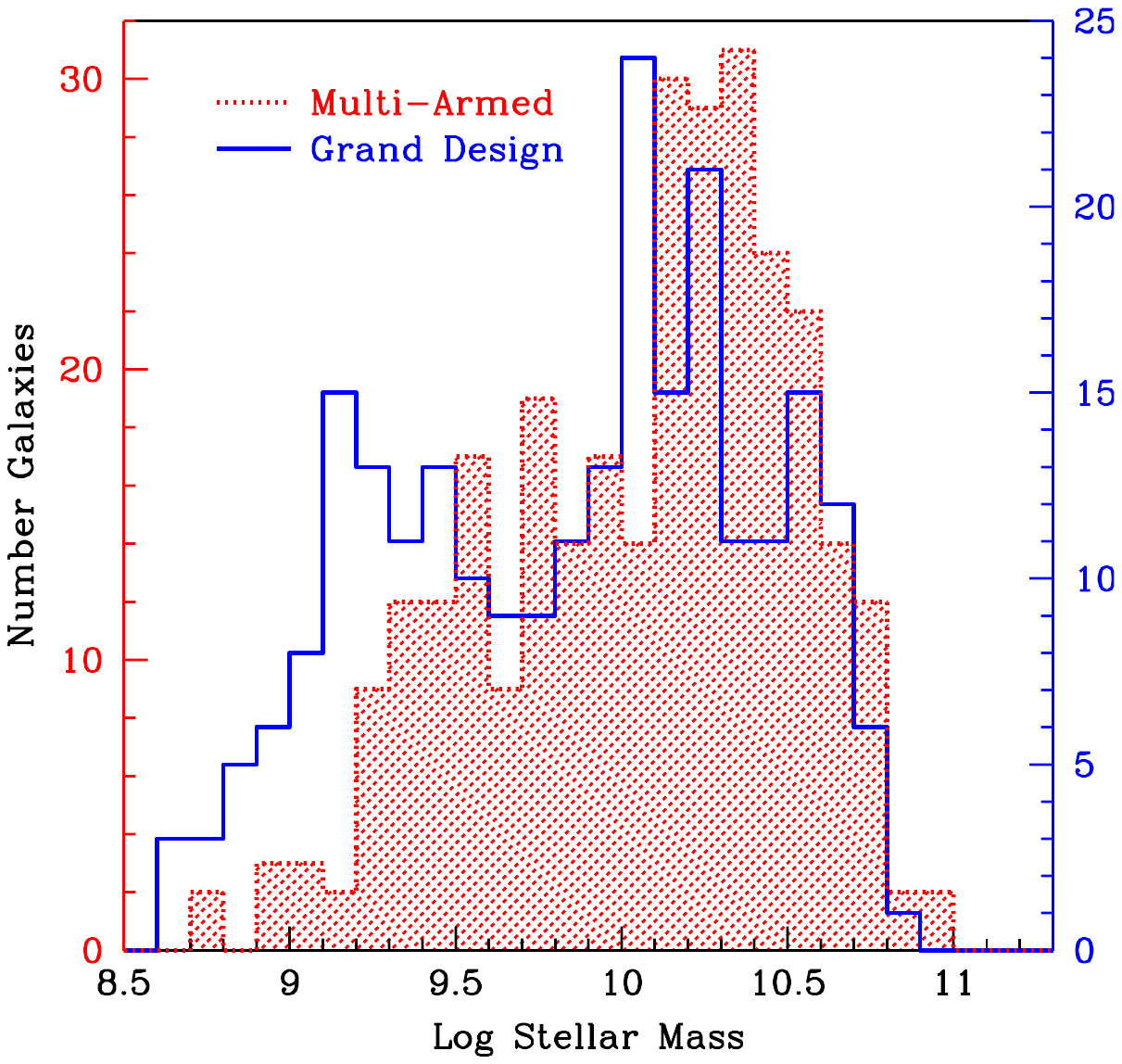}
\caption{ 
Histograms of redshifts (left panel)
and stellar masses (right panel)
for the final set of multi-armed galaxies 
(red hatched histogram)
and grand design spirals (blue histogram).
The left axis gives the number of galaxies for the multi-armed
sample, while the right axis gives the number of grand design
galaxies.
\label{fig:redshift_and_mass}}
\end{figure}

In Figure \ref{fig:redshift_and_mass}, we provide histograms
of the redshifts (left panel) and stellar masses
(right panel)
for our final sets of 245 grand design spirals
and 299 multi-armed galaxies.
The sample contains very few galaxies from 
0.01 $\le$ z $<$ 0.015,
followed by a broad peak 
from z = 0.02 $-$ 0.03, a drop at around z $\sim$ 0.035,
followed by a flat plateau which ends abruptly 
at our redshift limit of 0.05.
The drop-off at z $\sim$ 0.035 is due to the gap
between the primary and secondary samples as seen in 
Figure \ref{fig:redshift_vs_mass}.

In the following analysis, we 
use Kolmogorov-Smirnov (KS) and Anderson-Darling (AD) 
statistical tests to compare various properties of the multi-armed
galaxies vs.\ 
the grand design galaxies.
KS and AD tests 
give probabilities that two samples
came from the same parent sample, where a value greater 
than 0.05 
means that 
there is no significant 
difference 
between the two samples.
Our AD test implementation only provides a limit on the probability
if it is above $\ge$0.25 or below 0.001.
Both tests
find a 
significant difference between the redshifts of the two samples
(KS probability of 0.02; AD probability of 0.006).
The grand design galaxies have a slightly broader distribution
in z than the multi-armed galaxies, but a similar median z.
When the sample is divided into subsets by M*, a marginal
difference remains in the 9 $\le$ log (M*/M$_{\sun}$) $<$ 10 bin
(KS probability 0.055; AD probability 0.029), but no difference
is seen for galaxies in the range 10 $\le$ log (M*/M$_{\sun}$) $<$ 11. 

We see 
a clear difference in the stellar masses of our grand
design vs.\ our multi-armed galaxies.
The multi-armed galaxies
have a mass distribution that is skewed to higher masses
compared to
the grand design galaxies 
(Figure \ref{fig:redshift_and_mass}, right), 
a result confirmed by KS and AD tests 
(KS probability of 0.00024, AD probability $<$0.001).

\subsection{Arm Counts vs.\ Concentration } \label{sec:con}

\begin{figure}[ht!]
\plotone{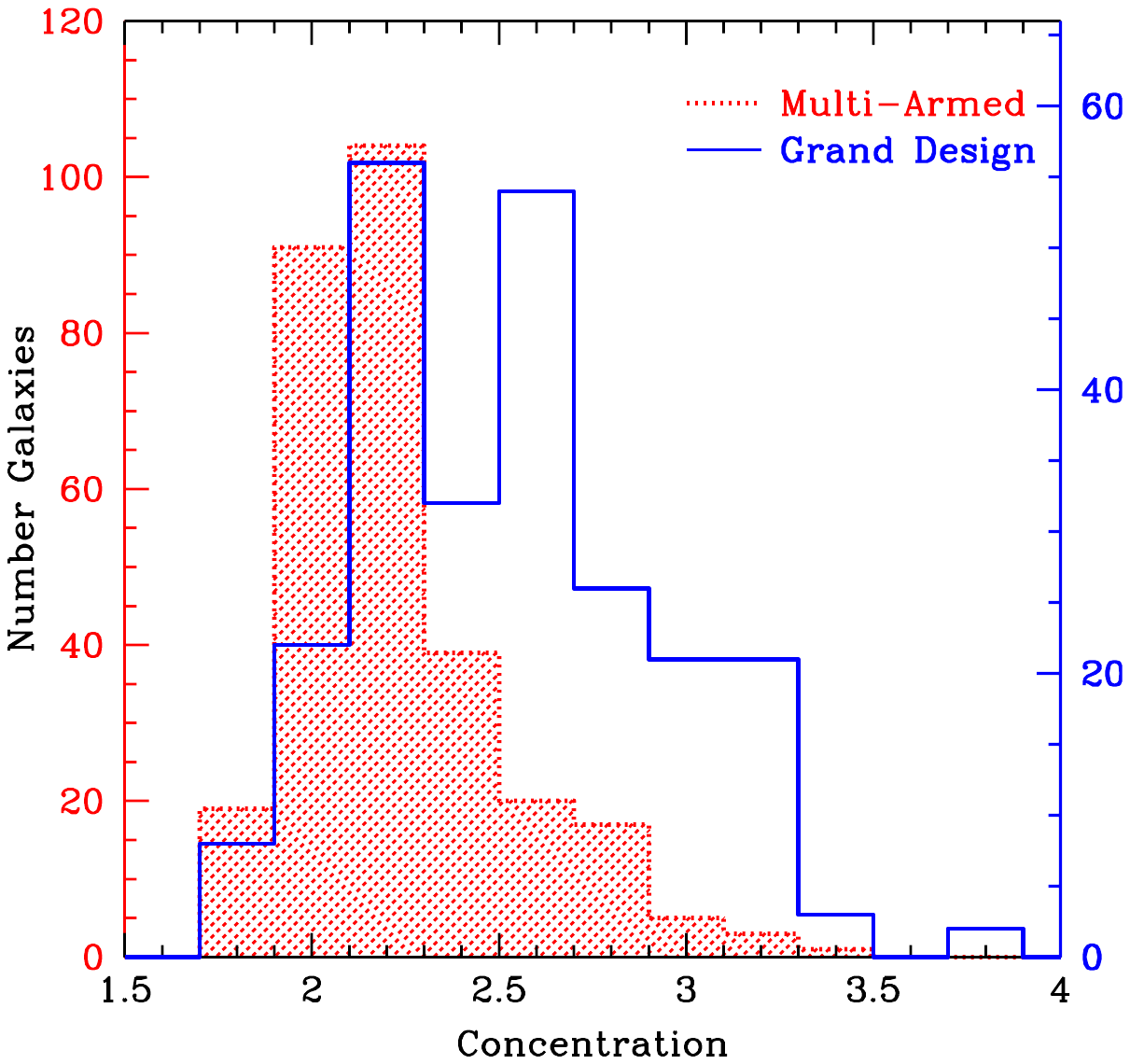}
\caption{ 
Histograms of the concentrations 
for 
our multi-armed (red hatched) vs.\ 
grand design galaxies (blue).
Concentration is defined as the ratio of R90/R50, where
R90 and R50 are the radii containing 90\% and 50\% of the SDSS r band
light, respectively, using elliptical Petrosian radii from the NSA.
The left axis gives the number of galaxies for the multi-armed
sample, while the right axis gives the number of grand design
galaxies.
\label{fig:NSAconcentration}}
\end{figure}

In this paper, we define concentration as 
the 
ratio of the 
radius containing 90\% of the SDSS r band light (R90) to the radius
containing 50\% of the light (R50)
\citep{2001AJ....122.1861S, 2004MNRAS.353..713K}, 
using the 
elliptical Petrosian radii from the NSA.
In 
Figure 
\ref{fig:NSAconcentration} we show histograms of
the concentrations 
for our sample of multi-armed galaxies 
vs.\ the grand design galaxies.
KS/AD test results for various stellar mass ranges are given in 
Table 
\ref{Ctab}.
Galaxies identified as multi-armed
have significantly lower concentrations
on average than the grand design galaxies.
This difference holds even when the galaxy is divided into subsets
by mass, except for 
the lowest mass bin (9.1 $\le$ log (M*/M$_{\sun}$) $<$ 9.4). 

\begin{figure}[ht!]
\plottwo{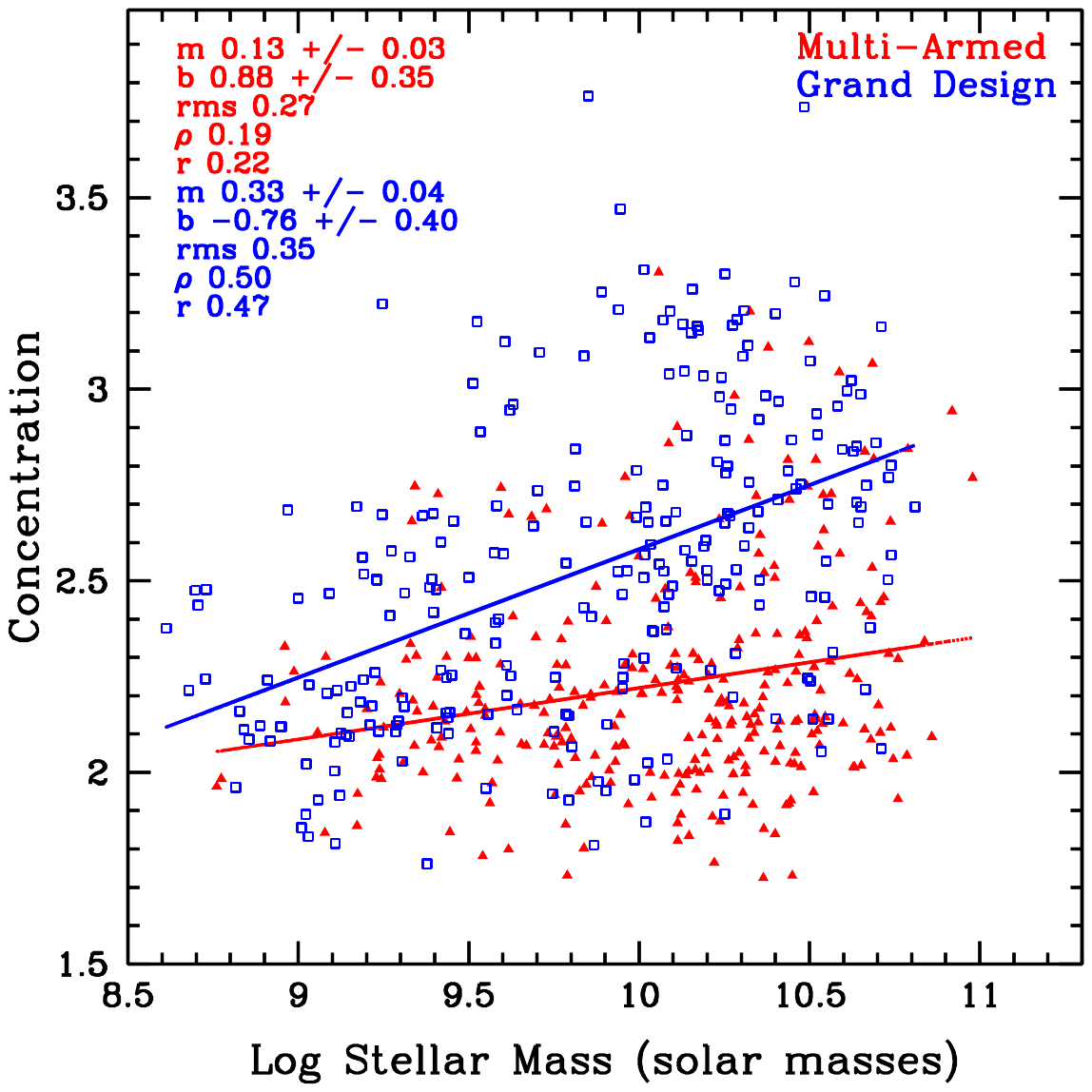}{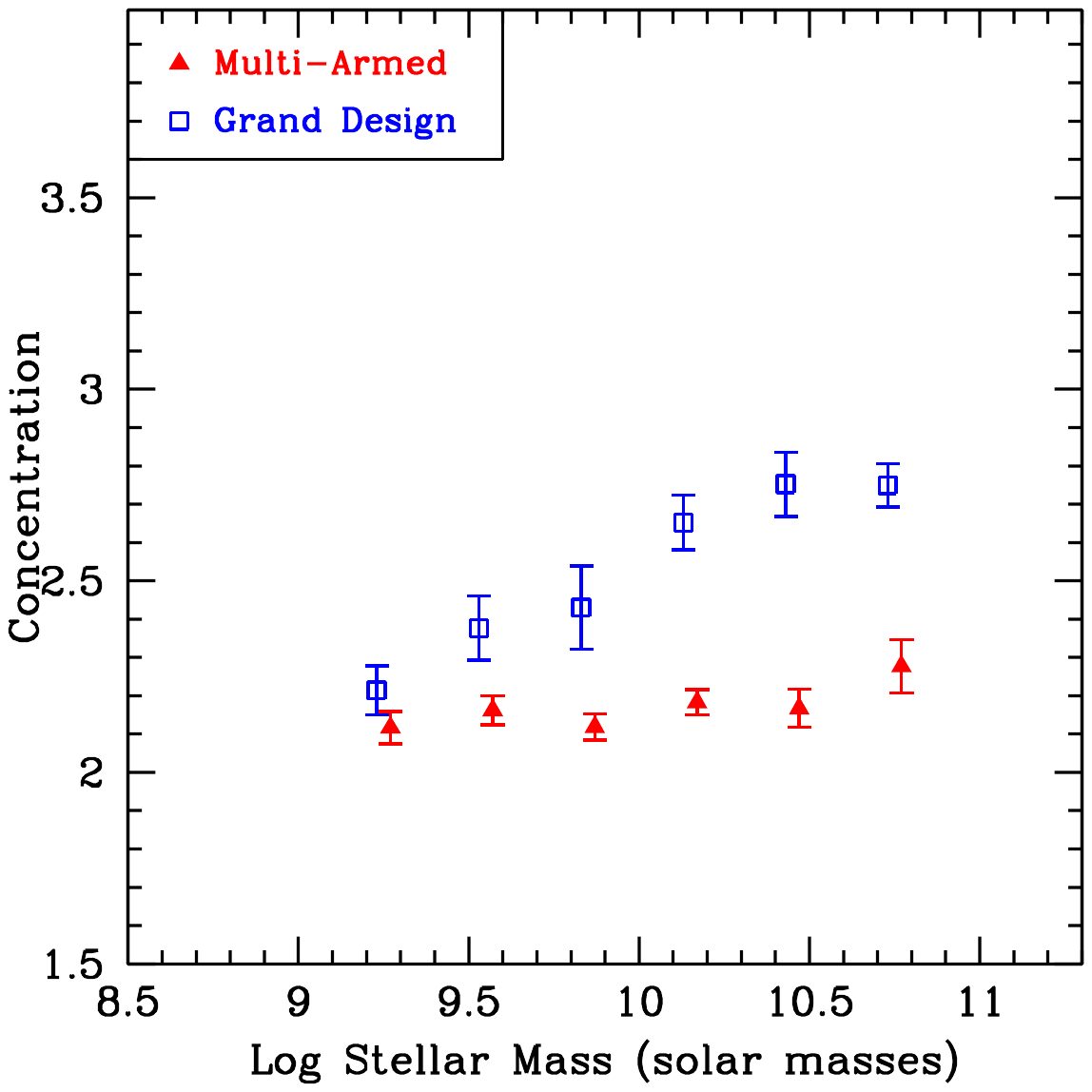}
\caption{ 
Left:
Plot of concentration vs.\ stellar mass for 
galaxies in our multi-armed galaxy sample (red filled triangles) 
and galaxies classified 
as grand design (blue open squares).  
Concentration is defined as the ratio of R90/R50, where
R90 and R50 are the radii containing 90\% and 50\% of the SDSS r band
light, respectively, using elliptical Petrosian radii from the NSA.
The red line is the best-fit for 
the multi-armed galaxies, while
the blue line is the best-fit for 
the grand design galaxies.
The best-fit slope (m), y-intercept (b), and rms
for the
two sets of galaxies are displayed in the corresponding color,
along with the 
Spearman ($\rho$) and Pearson (r) correlation
coefficient.
Right: plot of median concentration for the grand design
galaxies
(open blue squares) and the multi-armed galaxies (red filled triangles)
in bins in log M*.  
The error bars give the error in the median $\sim$ IQR/$\sqrt{N}$,
where IQR is the inter-quartile range and N is the number of datapoints
in that set.
The red and blue datapoints are slightly offset 
from the center of the mass bin to make the data easier to see.
Results of KS/AD tests comparing the grand design and the multi-armed
galaxies in each mass bin are given in Table \ref{Ctab}.
\label{fig:C_vs_mass}}
\end{figure}

Concentration is a function of stellar mass for spiral galaxies,
in that  
galaxies with higher masses tend to have larger concentrations.
This trend is shown in 
Figure 
\ref{fig:C_vs_mass}, 
where we plot concentration vs.\ stellar mass
for both the multi-armed and 
grand design galaxies.
A reasonably strong correlation is seen between mass and concentration for the 
grand design galaxies, while
the correlation is weaker for the
multi-armed galaxies.
Figure 
\ref{fig:C_vs_mass} shows a deficiency of high mass, low concentration
galaxies in the grand design sample, compared to the multi-armed sample.

In the right panel of 
Figure 
\ref{fig:C_vs_mass},
we plot the median concentrations 
from 
Table 
\ref{Ctab}
for grand design
galaxies 
(blue open squares) and multi-armed galaxies (filled red triangles)
within 0.3 dex bins in M*.   
Significant differences are present in all mass bins
except possibly the lowest mass bin.

\begin{deluxetable}{c|cc|c|ccc|ccc}
\tablecolumns{10}
\tablewidth{0pc}
\caption{Kolmogorov-Smirnov/Anderson-Darling Tests: Comparing Concentration for Multi-Armed vs. Grand Design Galaxies\label{Ctab}}
\tablehead{   
\colhead{Log}   
& \colhead{Number} 
& \colhead{Number}
& \colhead{KS/AD}
&\multicolumn{3}{c}{\underline{Multi-Armed}}
&\multicolumn{3}{c}{\underline{Grand Design}}
\\ 
\colhead{M*}
& \colhead{Multi-}
& \colhead{Grand }
& \colhead{Prob }
& \colhead{Median}
& \colhead{1st}
& \colhead{3rd}
& \colhead{Median}
& \colhead{1st}
& \colhead{3rd}
\\ 
\colhead{Range} 
& \colhead{Armed} 
& \colhead{Design }
& \colhead{ }
& \colhead{C} 
& \colhead{Quartile}
& \colhead{Quartile}
& \colhead{C}
& \colhead{Quartile}
& \colhead{Quartile}
\\ 
\colhead{(M$_{\sun}$)} 
& \colhead{Galaxies} 
& \colhead{Galaxies}
& \colhead{ }
& \colhead{} 
& \colhead{C}
& \colhead{C}
& \colhead{}
& \colhead{C}
& \colhead{C}
\\ 
}
\startdata
10 to 11 & 180 & 116 & {\bf 8$\times$10$^{-22}$}/{\bf $\le$0.001 } & 2.19 & 2.04 & 2.42 & 2.69 & 2.49 & 2.98 \\  
10 to 10.5 & 128 & 82 & {\bf 9$\times$10$^{-20}$}/{\bf $\le$0.001 } & 2.18 & 2.02 & 2.35 & 2.68 & 2.5 & 3.04 \\  
10.5 to 11 & 52 & 34 & {\bf 0.0001}/{\bf $\le$0.001 } & 2.28 & 2.11 & 2.59 & 2.7 & 2.46 & 2.91 \\  
9.1 to 9.4 & 23 & 39 & 0.075/{\bf 0.034 } & 2.12 & 2.02 & 2.22 & 2.21 & 2.11 & 2.5 \\   
9.4 to 9.7 & 38 & 32 & {\bf 0.01}/{\bf $\le$0.001 } & 2.16 & 2.07 & 2.3 & 2.38 & 2.18 & 2.66 \\  
9.7 to 10.0 & 50 & 33 & {\bf 0.00093}/{\bf $\le$0.001 } & 2.12 & 2.04 & 2.28 & 2.43 & 2.13 & 2.75 \\  
10 to 10.3 & 73 & 60 & {\bf 6$\times$10$^{-14}$}/{\bf $\le$0.001 } & 2.18 & 2 & 2.28 & 2.65 & 2.48 & 3.03 \\  
10.3 to 10.6 & 77 & 37 & {\bf 2$\times$10$^{-7}$}/{\bf $\le$0.001 } & 2.17 & 2.07 & 2.51 & 2.75 & 2.46 & 2.97 \\  
10.6 to 10.9 & 28 & 19 & {\bf 0.00012}/{\bf $\le$0.001 } & 2.28 & 2.09 & 2.46 & 2.75 & 2.61 & 2.86 \\  
\enddata
\end{deluxetable}

\subsection{Arm Counts vs.\ Physical Size} \label{sec:size}

\begin{figure}[ht!]
\plottwo{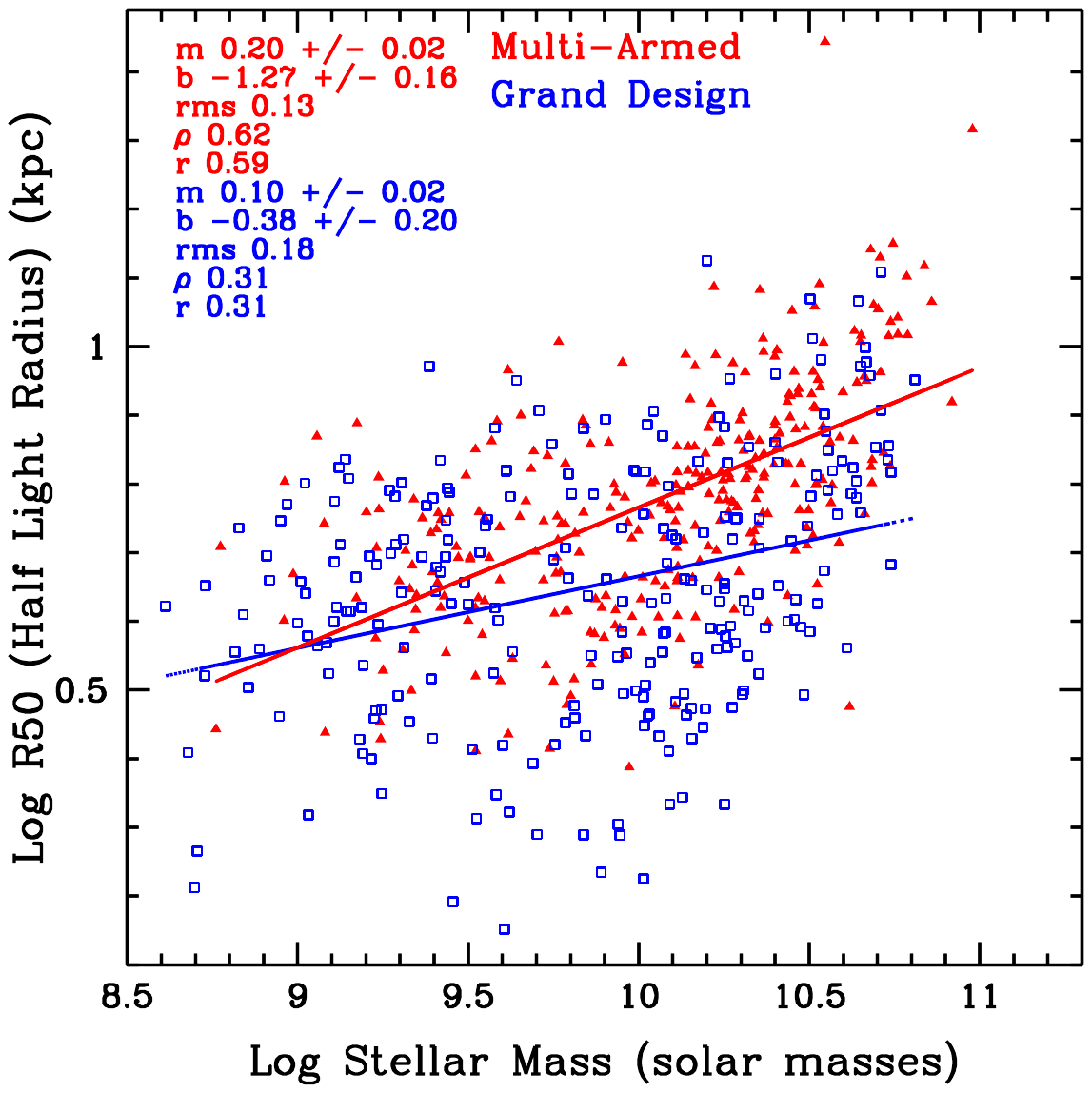}{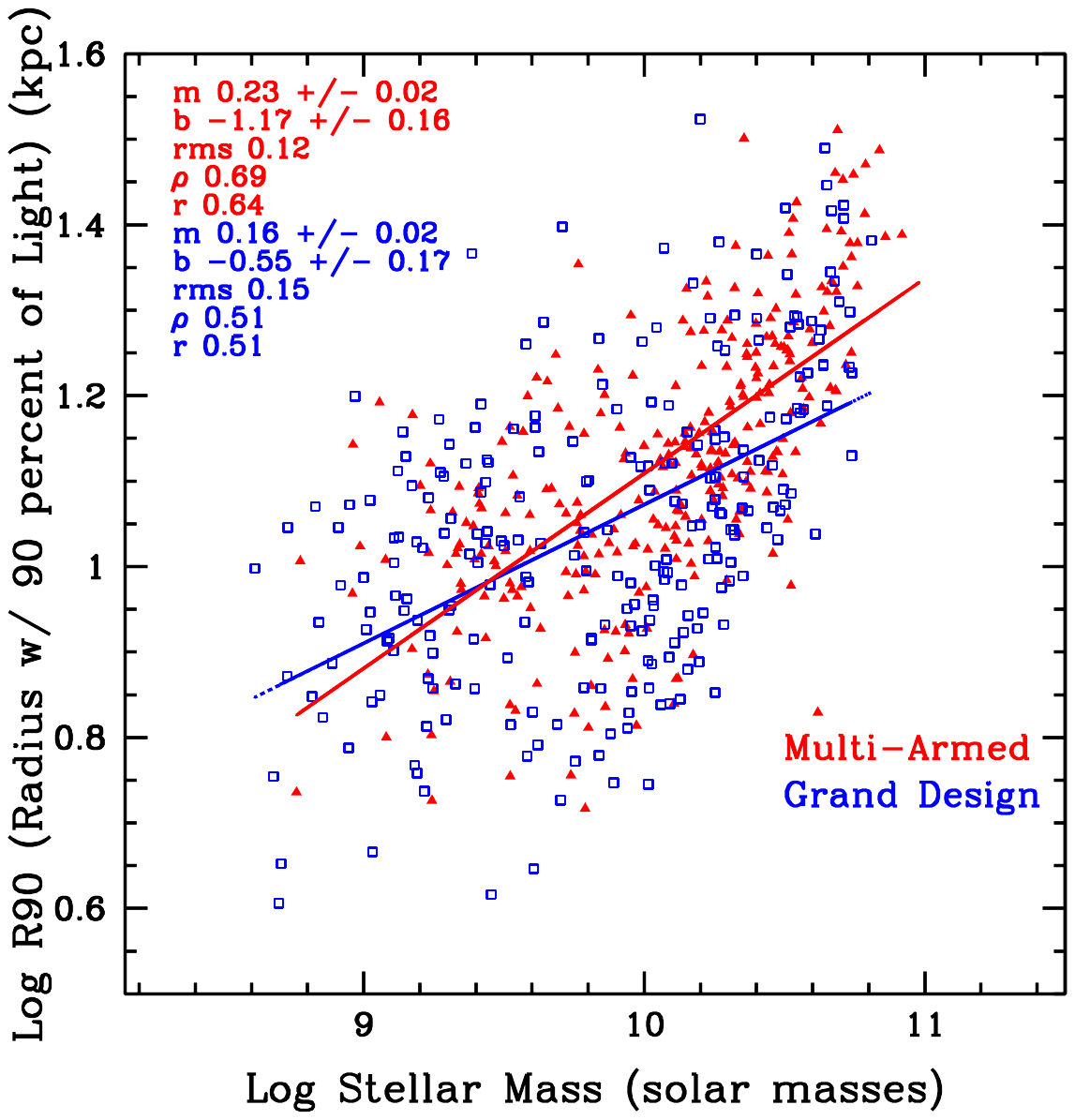}
\caption{ 
Left:
Plot of the log of the 
SDSS r band R50 radius
(half light radius)
from the NSA
vs.\ the log of the stellar mass,
for
our multi-armed (red filled triangles)
and grand design galaxies (blue open squares).
Right: Plot of R90, the radius containing 90\% of the SDSS r band light,
vs.\ the stellar mass
for our sample of multi-armed galaxies (red filled triangles)
and the grand design galaxies (blue filled squares).
The red lines are the best-fit linear relations 
for the multi-armed galaxies, while 
the blue lines 
are the best-fit lines for the grand design galaxies.
The parameters
of the best-fit lines and the 
Spearman ($\rho$) and Pearson (r) correlation
coefficients are overlaid in the corresponding color.
\label{fig:Size1}}
\end{figure}

In Figure \ref{fig:Size1}, we plot two different estimates of galaxy
size vs.\ stellar mass, for our two classes of spiral
galaxies. The left panel shows
the log of the SDSS r band R50 radius (half light radius) from the NSA
vs.\ log M*, while the right panel displays log R90 vs.\ log M*.
For R50, the
slope of the best-fit line is significantly steeper for the
multi-armed galaxies
compared to that for the grand design galaxies
(left panel, Figure \ref{fig:Size1}).
Above  
log (M*/M$_{\sun}$) $>$ 10, there is a clear offset between the two
classes of spiral galaxies, with multi-armed galaxies having 
larger R50 for a given stellar mass.
The correlation between log R50 and log M* is stronger for the 
multi-armed galaxies.   For R90, the difference between
grand design and multi-armed galaxies is smaller; the slope of the
best-fit line for multi-armed 
galaxies is only marginally larger than that of grand design galaxies.
In 
Figure \ref{fig:Size1}, 
there is a hint of a bend in the relation near log (M*/M$_{\sun}$)  = 10, with
a steeper slope at higher masses.

These plots show that the total extent of stellar disks in the two
classes of spirals are similar for a given stellar mass; 
the differences in concentration (R90/R50) between the two classes
of galaxies
are associated with grand design galaxies having smaller R50 values for a given
R90 size than multi-armed galaxies, meaning 
they have proportionally more light in the center.

\subsection{Arm Count vs.\ Hubble Type}

\begin{figure}[ht!]
\plottwo{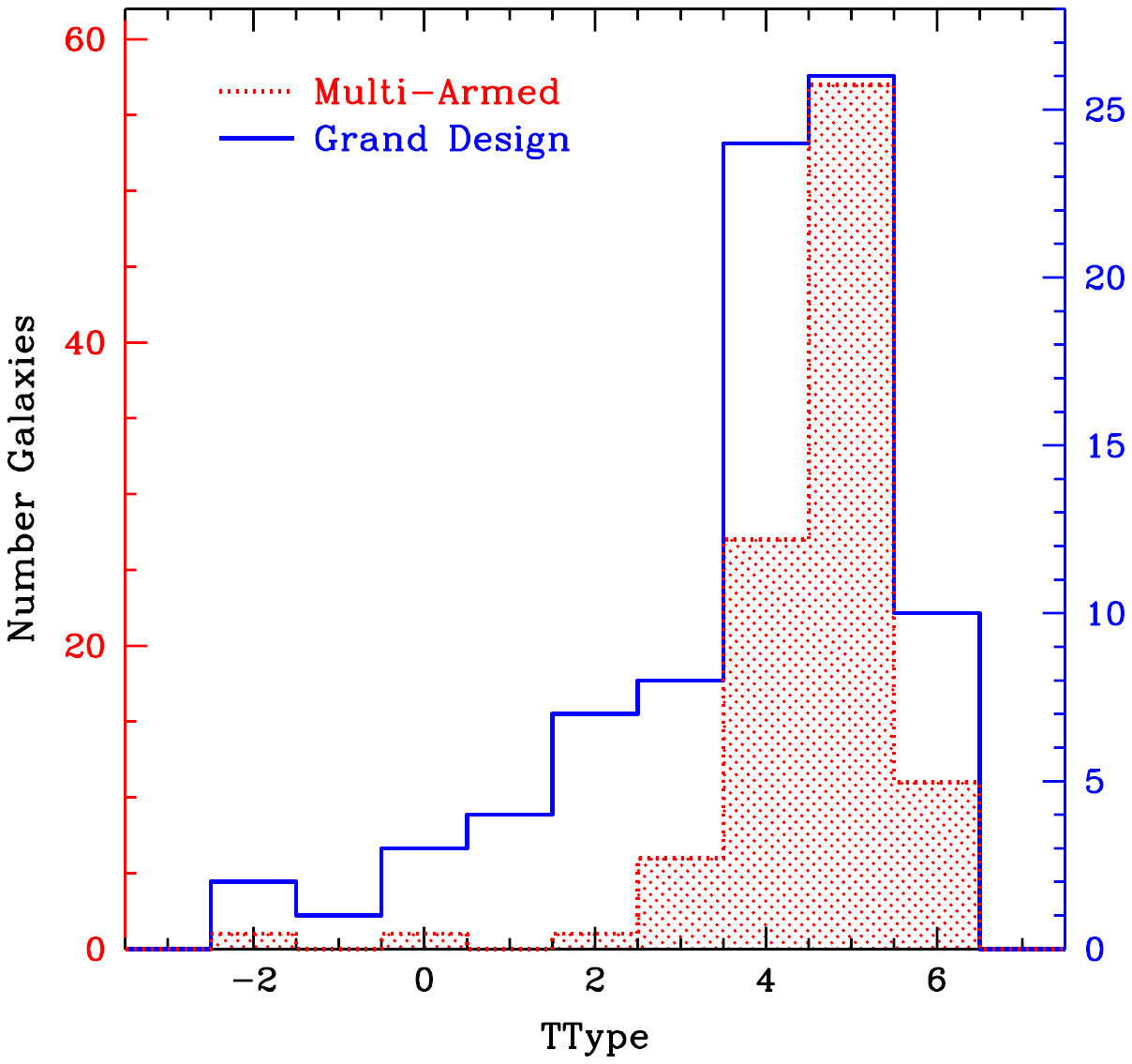}{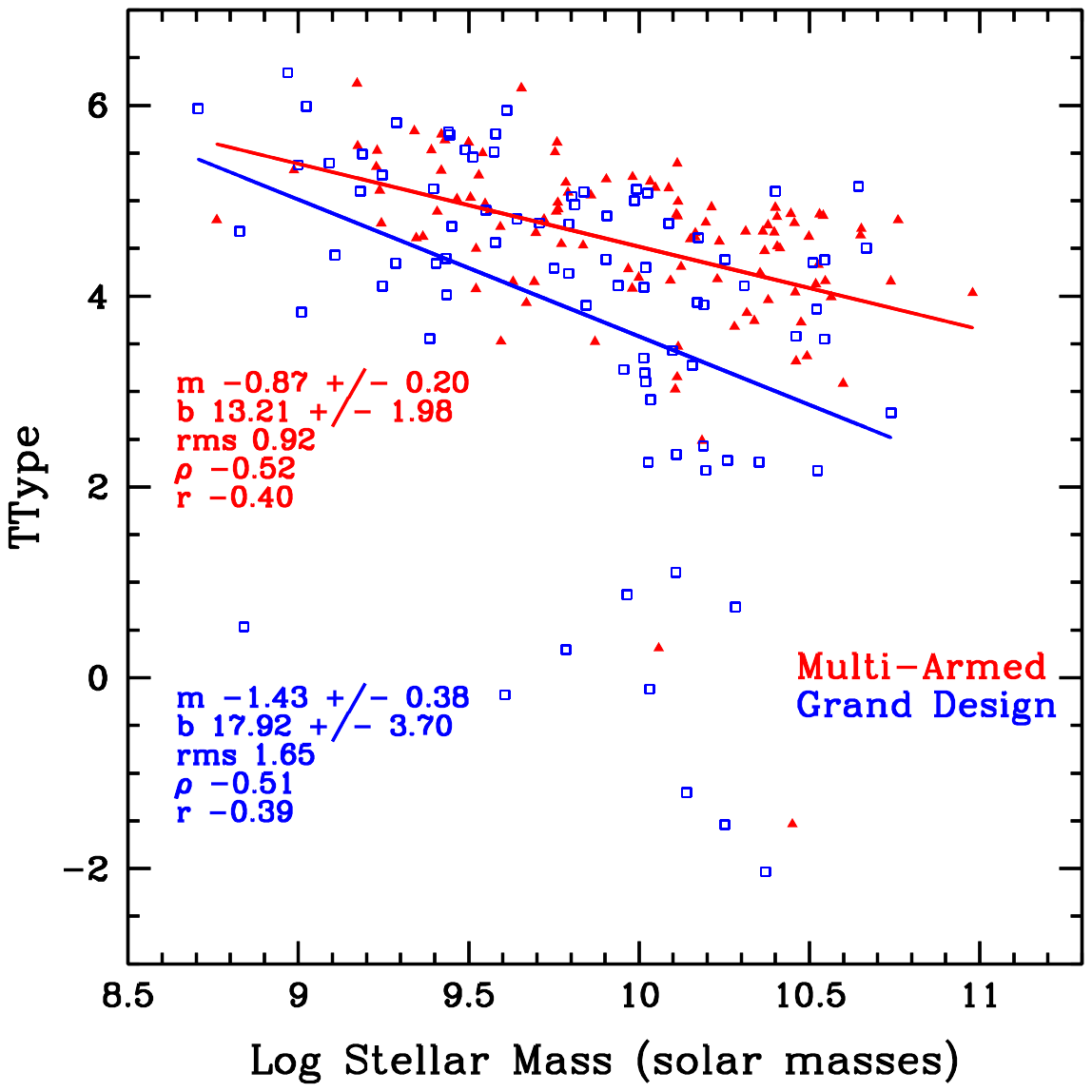}
\caption{ 
Histograms of the 
\citet{2018MNRAS.476.3661D}
Hubble Type (TType) 
for
our multi-armed (red hatched)
and grand design galaxies (blue).
The left axis gives the number of galaxies for the multi-armed
sample, while the right axis gives the number of grand design
galaxies. 
Right: Plot of the 
\citet{2018MNRAS.476.3661D}
Hubble Type (TType) 
vs.\ log stellar mass
for
our 
multi-armed galaxies (red filled triangles)
and grand design galaxies (blue open squares).
The best-fit line for the multi-armed galaxies
is plotted in red, while the best-fit 
line for the grand design galaxies is given in blue.
The parameters of the best-fit line are
given in the corresponding color.
\label{fig:DLHubType}}
\end{figure}

For many galaxies in the NSA, 
\citet{2018MNRAS.476.3661D}
used machine learning with Convolutional Neural Networks
to estimate their TType, 
where 
TTypes are Hubble types coded into numerical values.
Their code was trained
using the 
\citet{2010ApJS..186..427N}
by-eye classifications of Hubble Types
as well as GZ2 classifications 
from \citet{2013MNRAS.435.2835W}.
In the left panel of 
Figure \ref{fig:DLHubType},
we provide histograms of the 
\citet{2018MNRAS.476.3661D}
TTypes
for our final samples of multi-armed and grand design 
galaxies.
The distribution of TTypes for the grand design galaxies
is skewed toward
earlier-type galaxies
compared to the multi-armed galaxies.  
KS and AD tests for the full mass range confirm that there is a significant
difference in TTypes between the two samples
(KS probability 0.0047, AD probability 0.002).

Like concentration, TType also depends upon mass.
In the right panel of Figure \ref{fig:DLHubType}, we plot 
TType vs.\ stellar mass for
our multi-armed and grand design galaxies. 
For both arm classes, there is a strong correlation, in that lower
mass galaxies have later TTypes.   
For a given stellar mass,
the grand design galaxies tend to have earlier types, however,
the slopes of the best-fit relations are consistent within the uncertainties. 
When the sample is sub-divided by stellar mass into 1 dex bins, 
a difference in TType is seen
in the 
10 $\le$ log (M*/M$_{\sun}$) $<$ 11 
bin 
(KS probability 0.0023, AD probability $\le$0.001)
but not in
9 $\le$ log (M*/M$_{\sun}$) $<$ 10. 
Sub-dividing further
into 0.3 dex bins, only the 10 $\le$ { log (M*/M$_{\sun}$) $<$ 10.3 }
bin 
shows a
significance difference 
(KS probability 0.0023, AD probability $\le$0.001), with the  
median TType
for grand design galaxies being less than for multi-armed galaxies.

\subsection{Arm Counts vs.\ Central Surface Mass Density} \label{sec:Sigma}

\begin{figure}[ht!]
\plottwo{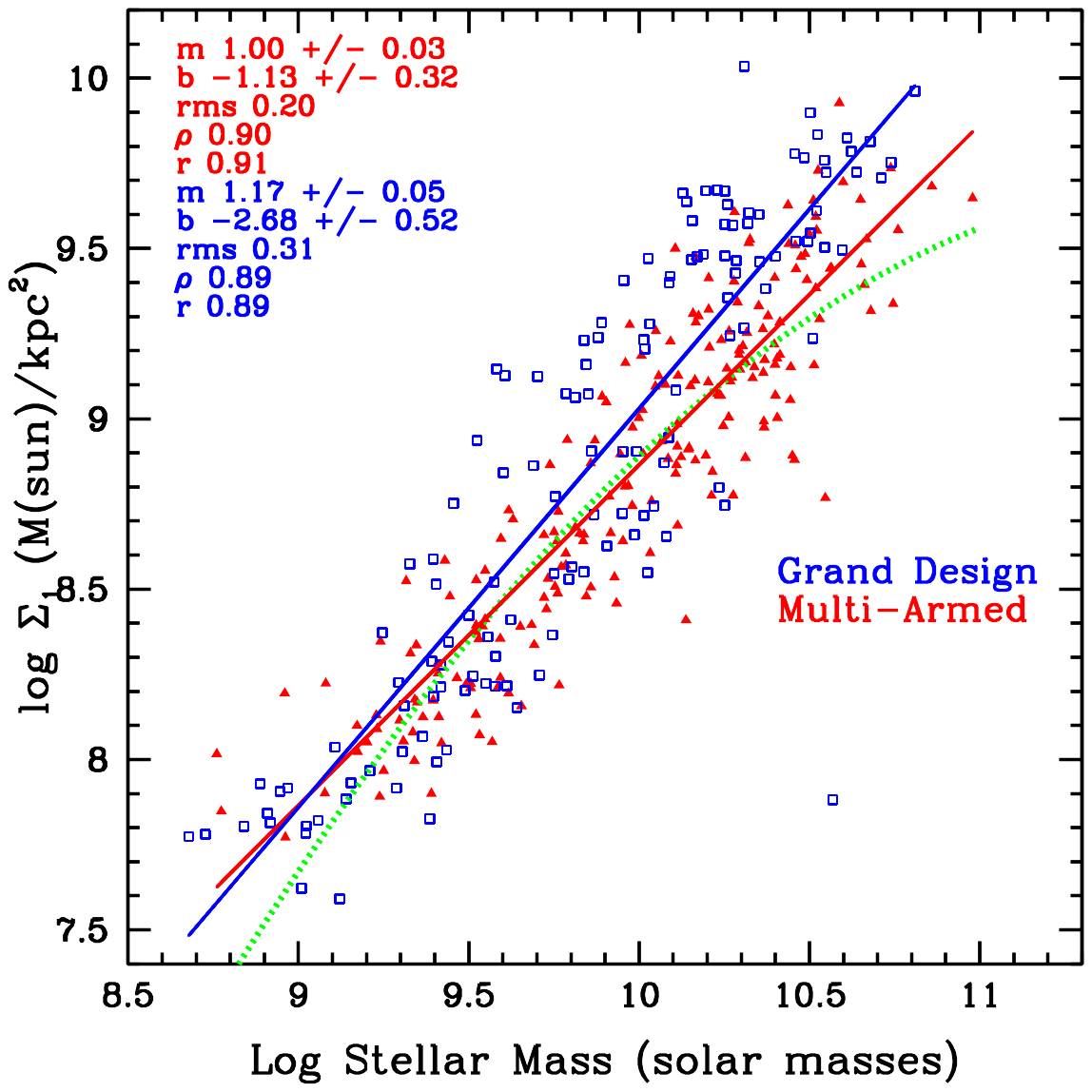}{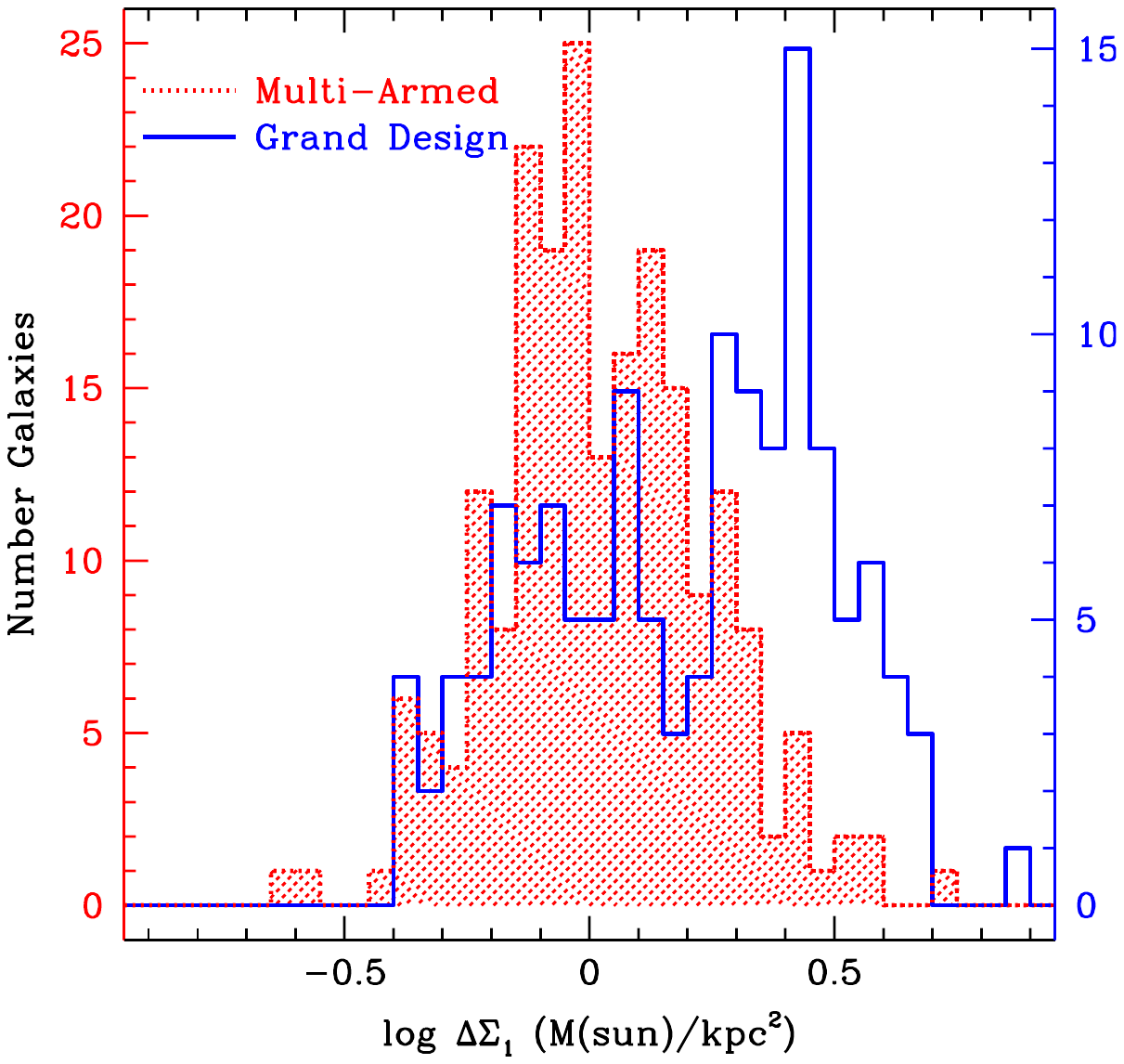}
\caption{ 
left: Plot of log $\Sigma$$_1$ vs.\ log M* for
the 
multi-armed (red filled triangles) and grand design (blue open squares) galaxies.
The red line is the best-fit line for the multi-armed
galaxies; the blue line is the best-fit line
for the grand design galaxies.  The parameters of the fits are given
in the corresponding colors.
The green dotted curve is the 
\citet{2020MNRAS.493.1686L}
dividing line
between galaxies with classical bulges (above the curve)
and pseudo-bulges (below).
Right: 
log $\Delta$$\Sigma$$_1$ for multi-armed (red hatched)
and grand design (blue) galaxies, 
where 
log $\Delta$$\Sigma$$_1$ is defined by
\citet{2020MNRAS.493.1686L} as the offset from the green curve in the 
left panel.
Galaxies that lie to the right of zero in these histograms lie
above the green curve in the plot on the left. 
The left axis gives the number of galaxies for the multi-armed
sample, while the right axis gives the number of grand design
galaxies.
\label{fig:Sigma1}}
\end{figure}

Measurements of the surface mass density within the central 1 kpc radius
($\Sigma$$_1$)
are available 
from \citet{2020MNRAS.493.1686L}
for 209 of our multi-armed galaxies and 135 of our grand design galaxies.
In the left panel of 
Figure \ref{fig:Sigma1}, 
we plot log $\Sigma$$_1$ against log M* for both classes of spirals.
In both cases, 
$\Sigma$$_1$ is correlated with M*.   However,
grand design galaxies show a steeper slope for the best-fit line,
and 
an offset is seen between the two arm classes at high masses;
for a given M*, grand design galaxies have higher $\Sigma$$_1$ on average.  

\citet{2020MNRAS.493.1686L} 
concluded that $\Sigma$$_1$ is related to 
bulge type; galaxies with bulges classified as pseudo-bulges
have lower $\Sigma$$_1$ for a given M* than galaxies with 
classical bulges. A pseudo-bulge is a disk-like structure
in the center of a spiral galaxy that masquerades as a bulge
and was produced by secular processes, while  
a classical bulge resembles an elliptical embedded within a disk galaxy
\citep{2004ARA&A..42..603K}.
In the left panel of Figure \ref{fig:Sigma1}, 
we overlay the 
dividing curve between the two classes of bulges
as determined by 
\citet{2020MNRAS.493.1686L}.   
Proportionally more grand design 
galaxies lie above the 
\citet{2020MNRAS.493.1686L} curve, relative to the multi-armed galaxies.

In the right panel of Figure 
\ref{fig:Sigma1},
we provide histograms of
log
$\Delta$$\Sigma$$_1$, 
where
log
$\Delta$$\Sigma$$_1$
is defined 
by 
\citet{2020MNRAS.493.1686L}
as the 
offset in 
log $\Sigma$$_1$ from
the green line in the left panel of 
Figure \ref{fig:Sigma1}.
With this definition, 
according to 
\citet{2020MNRAS.493.1686L}
log
$\Delta$$\Sigma$$_1$ is positive for 
galaxies with classical bulges, and negative for galaxies with pseudo-bulges.
Figure \ref{fig:Sigma1}
shows that a larger fraction of grand design galaxies have
log $\Delta$$\Sigma$$_1$ $>$ 0 compared to multi-armed galaxies.
A KS test gives a probability of only 3 $\times$ 10$^{-10}$
that the two distributions of 
$\Delta$$\Sigma$$_1$ come from the same parent population. 
Using a cutoff of log $\Delta$$\Sigma$$_1$ = 0 as the dividing line
between the two bulge types, 70\% of the grand design
galaxies host classical bulges, but only 50\% of the multi-armed galaxies.
This result is discussed further in Sections
\ref{sec:pb_cb} and 
\ref{sec:models}.

\subsection{Arm Count vs.\ Specific Star Formation Rate} \label{sec:sSFR}

\begin{figure}[ht!]
\plottwo{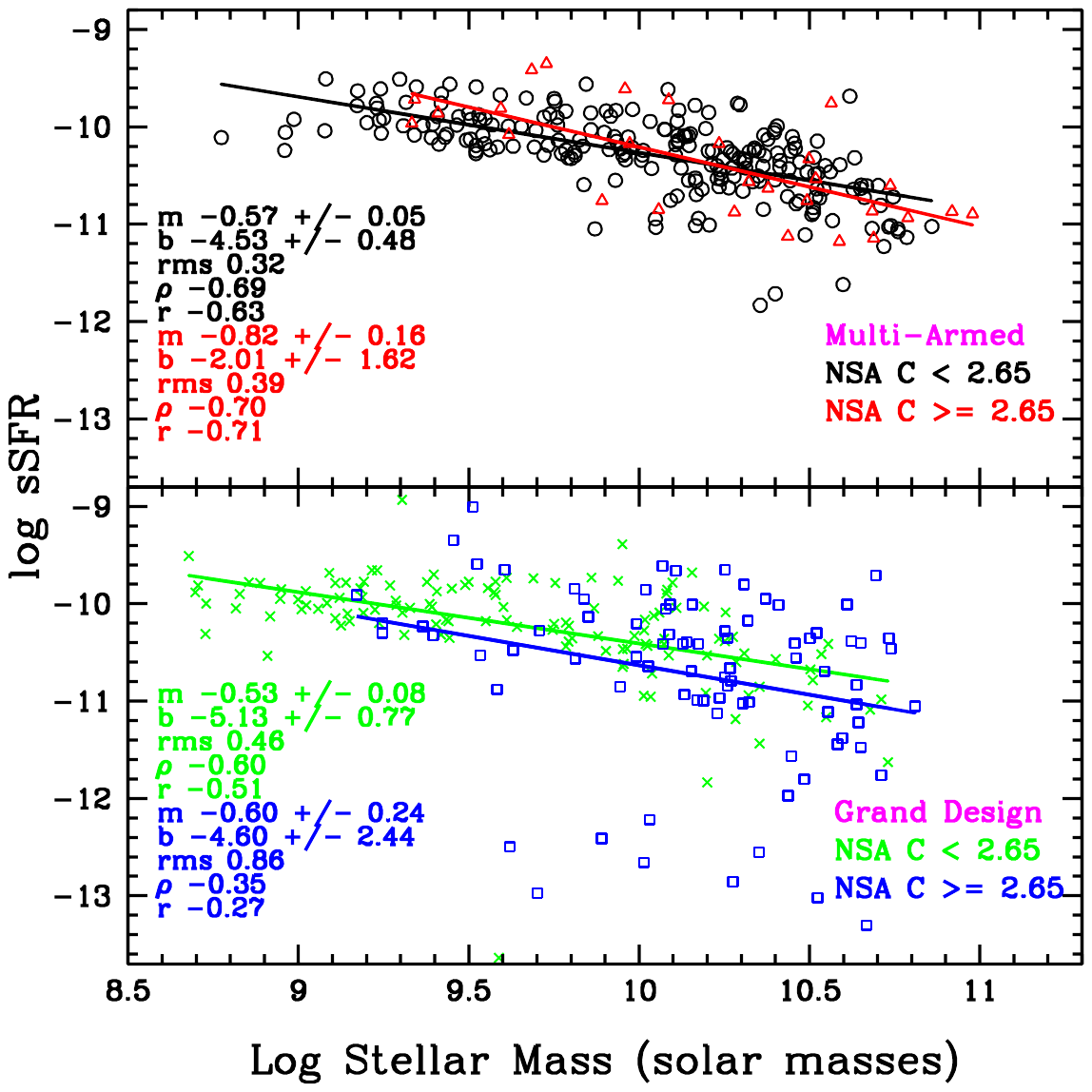}{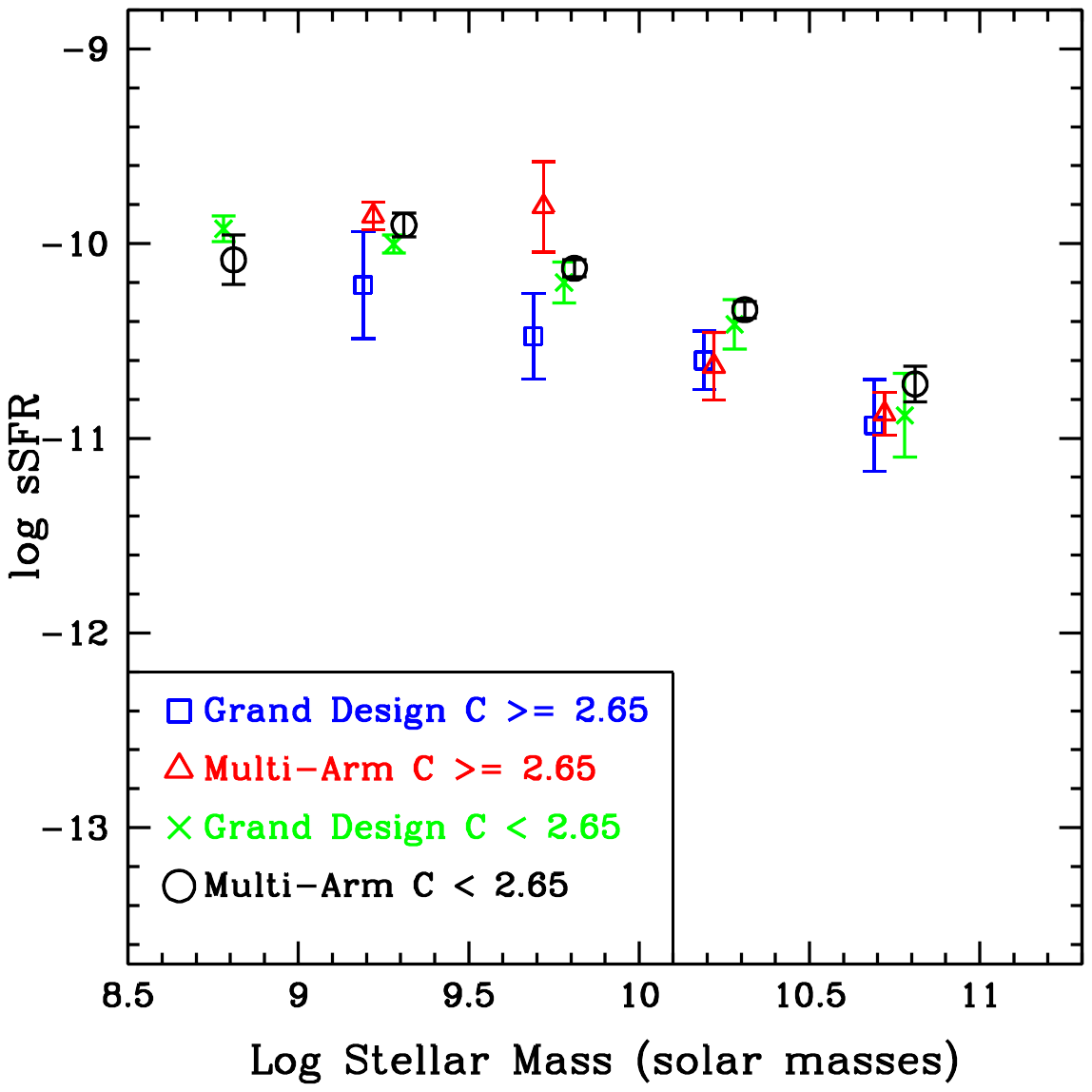}
\caption{ 
Plots of the log of the sSFR vs.\ log M*. 
The left panels show the positions of
individual galaxies, with grand design galaxies
in the top panel and multi-armed 
galaxies
in the bottom panel.  
In the panels on the left,  
the best-fit line for each subset and its
parameters are shown in the corresponding color.
The right panel gives 
the median sSFR for each
subset as a function of log M*, within 0.3 dex bins in log M*.
The sample is split into four subsets:
blue open squares: grand design, NSA C $\ge$ 2.65;
green crosses: grand design, NSA C $<$ 2.65;
red open triangles: 3-armed, NSA C $\ge$ 2.65;
black open circles: 3-armed, NSA C $<$ 2.65.
In the right panel, 
the four subsets for a given log M* are slightly 
shifted along the
x-axis for ease in viewing.
The error bars in the right panel give the error in the median $\sim$ IQR/$\sqrt{N}$,
where IQR is the inter-quartile range and N is the number of datapoints
in that set.
Only for the low C 9.5 $\le$ log (M*/M$_{\sun}$) $<$ 10 bin is there a marginally significant
difference between the grand design and multi-armed galaxies, and only for the KS test.
\label{fig:mainseq}}
\end{figure}

In Figure \ref{fig:mainseq},
we compare 
the 
sSFRs 
for our multi-armed and grand design
galaxies.
For this analysis, 
we use sSFRs from the GALEX-Sloan-Wise Legacy Catalog version 2 (GSWLC-2)
\citep{2016ApJS..227....2S, 2018ApJ...859...11S}.
Over the whole sample we see no significant difference in sSFR for the two
classes (KS/AD probabilities 0.21/0.057).
However, the sSFR of spirals is a function of stellar mass and concentration
\citep{2022AJ....164..146S}.
To determine how sSFR depends upon
the number of arms, it is important to separate out the effect of M* and C.
In the left panels of Figure \ref{fig:mainseq}, 
we plot log sSFR vs.\ log M*
for the two classes of galaxies (top panel:
grand design; bottom panel: multi-armed).  
Galaxies with lower stellar masses tend to have higher
sSFRs.
In the left panels
of Figure \ref{fig:mainseq}, 
the galaxies are further divided into
two sets based on concentration.  
The best-fit linear relations for these subsets are shown.  
For the grand design galaxies, galaxies with larger concentrations
have lower sSFRs on average than galaxies
with smaller concentrations.
For multi-armed galaxies, the best-fit log sSFR vs.\ log M*
relation for high C and low C galaxies are reasonably consistent.

In the right panel of Figure \ref{fig:mainseq},
we plot the median sSFR of grand design 
and multi-armed galaxies in 0.5 dex wide
bins
in M*, after dividing into two bins in concentration.
We ran KS and AD tests on these subsets.
The only bin for which a marginally significant difference is seen
is the 
9.5 $\le$ log (M*/M$_{\sun}$) $<$ 10, 1.55 $\le$ C $<$ 2.65 bin, with the
multi-armed galaxies having a higher median sSFR.  
When we account for differences in concentration 
between grand design
and multi-armed galaxies, we cannot rule out that 
sSFR is independent
of the number of arms in a galaxy. 

We repeated this analysis using log $\Delta$$\Sigma$$_1$ rather than
concentration.  We divided the galaxies into five bins in M* and 
two bins in log $\Delta$$\Sigma$$_1$, and compared the sSFRs of the
multi-armed and 
grand design galaxies in those bins.  
Only in the 9.0 $\le$ log (M*/M$_{\sun}$) 
$<$ 9.5, log $\Delta$$\Sigma$$_1$ $<$ 0 bin did we find 
that 
the multi-armed galaxies have significantly higher sSFRs than the  
grand designs (KS probability = 0.006).  However, there are
very few galaxies in that bin (14 of each kind), so this result
is uncertain.
With the current data,
we cannot rule out that sSFR is independent of the number of arms,
except indirectly via the correlation of sSFR
with concentration, log $\Delta$$\Sigma$$_1$,
and mass.

\section{Comparisons With Published Arm Counts} \label{sec:published}

\subsection{Comparison with \citet{2020ApJ...900..150Y} Data} \label{sec:YuHo}

\begin{figure}[ht!]
\plottwo{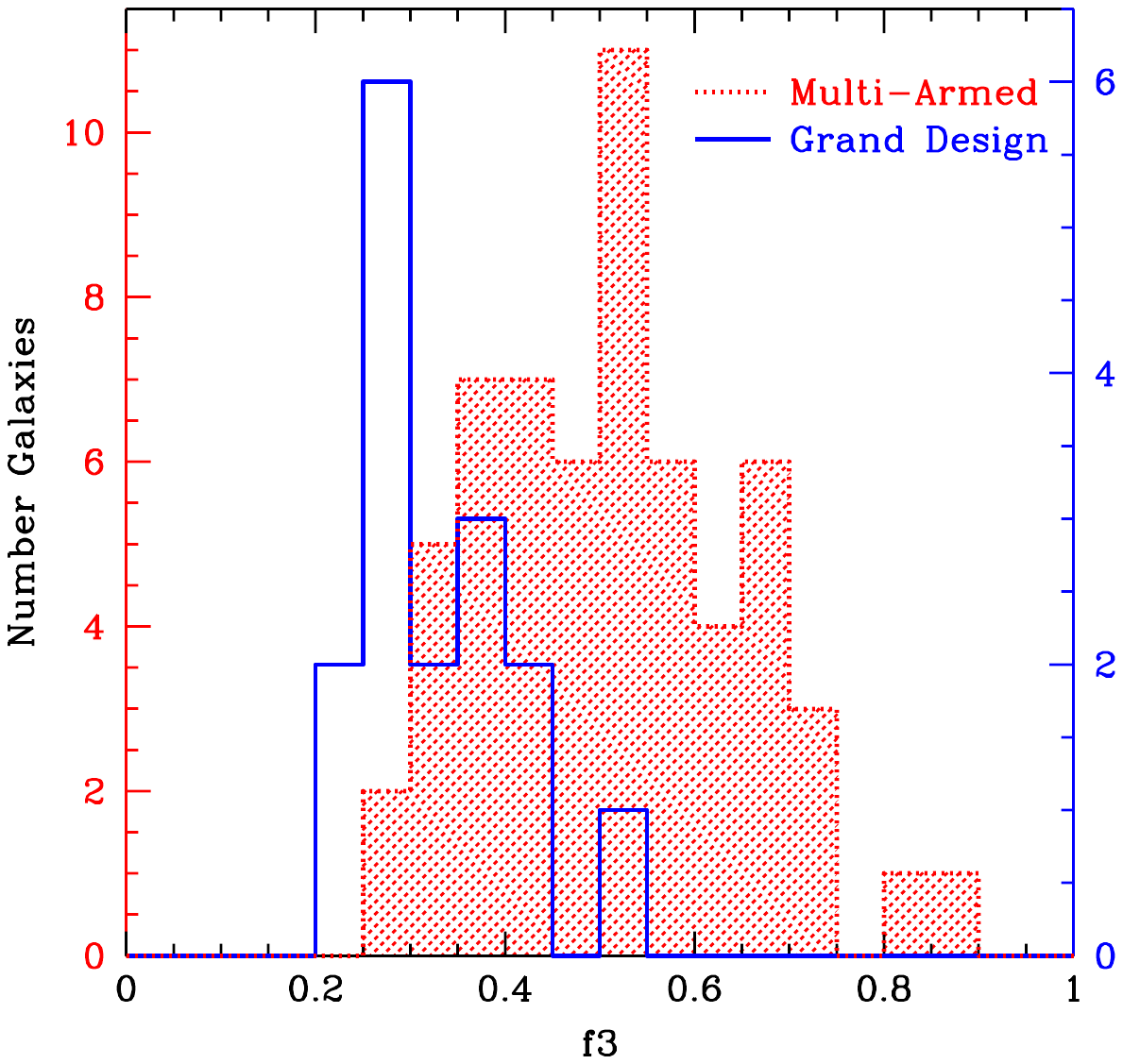}{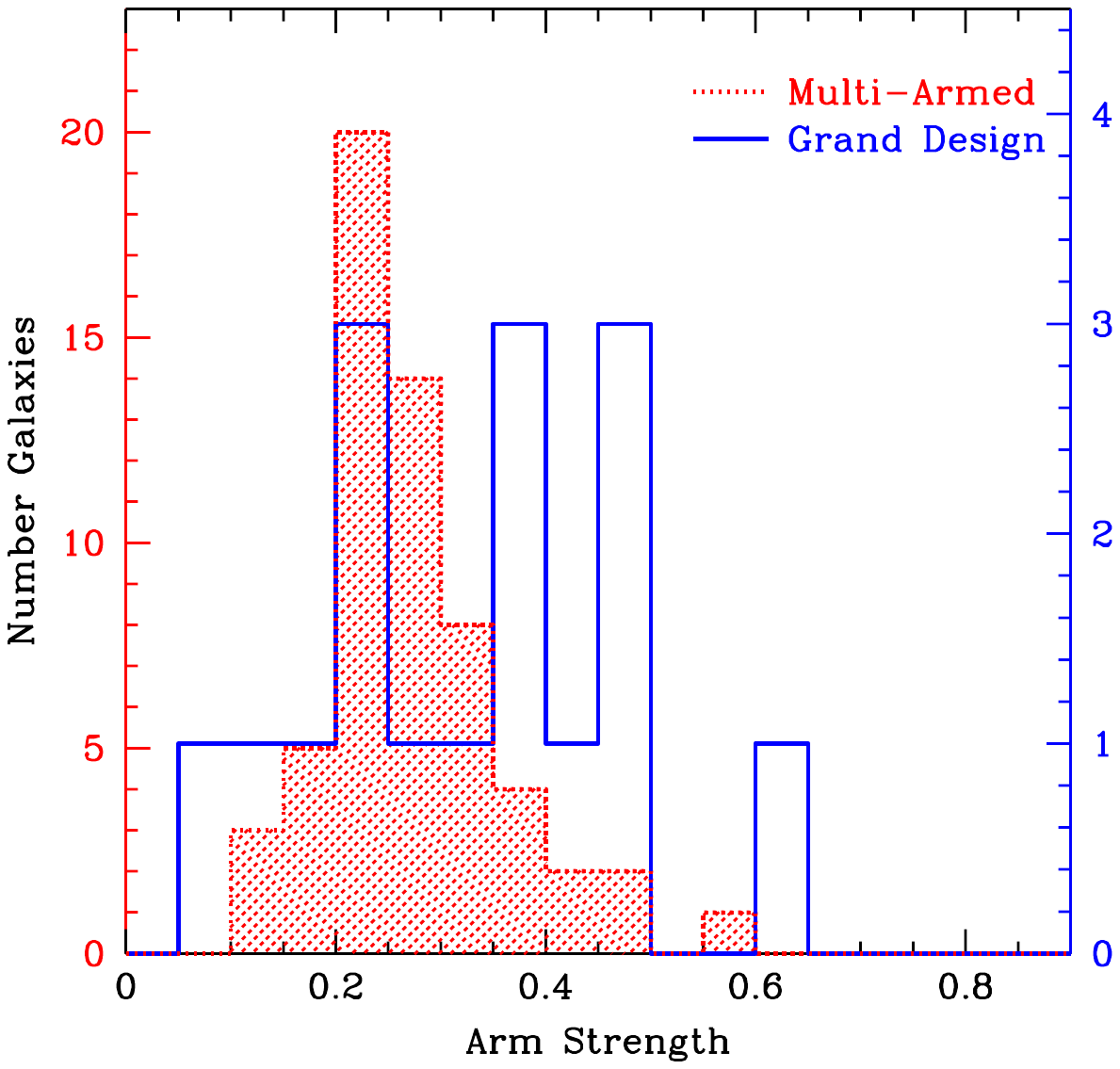}
\caption{ 
Left panel: Histograms of the \citet{2020ApJ...900..150Y} f3 parameter 
for our multi-armed galaxies (red hatched) and grand design galaxies (blue).
Right panel: Histograms of the \citet{2020ApJ...900..150Y}
arm strengths for
our multi-armed (red) and grand design (blue) galaxies.
\label{fig:f3}}
\end{figure}

In this section, we compare our arm classes with 
arm parameters from the
\citet{2020ApJ...900..150Y}
survey, for galaxies in common between the two samples.
Of the 1002 galaxies that are in both the GZ-3D
and the \citet{2020ApJ...900..150Y} samples, 
56 are in our final sample of multi-armed 
galaxies and 16 are in our final sample of 
grand design galaxies.
In 
the left panel of 
Figure \ref{fig:f3},
we compare 
histograms
of f3 for galaxies 
in our multi-armed sample 
vs.\ galaxies in the grand design sample.
Multi-armed galaxies have significantly
larger f3 values than grand design galaxies
(KS probability 4 $\times$ 10$^{-5}$).
The smaller f3 values for the grand design galaxies are 
consistent with expectations, since grand design galaxies
are expected to be dominated by the m=2 Fourier component, and thus
have low f3 values.
This agreement supports the idea that our method of separating
grand design and multi-armed galaxies is reliable.

\begin{figure}[ht!]
\plottwo{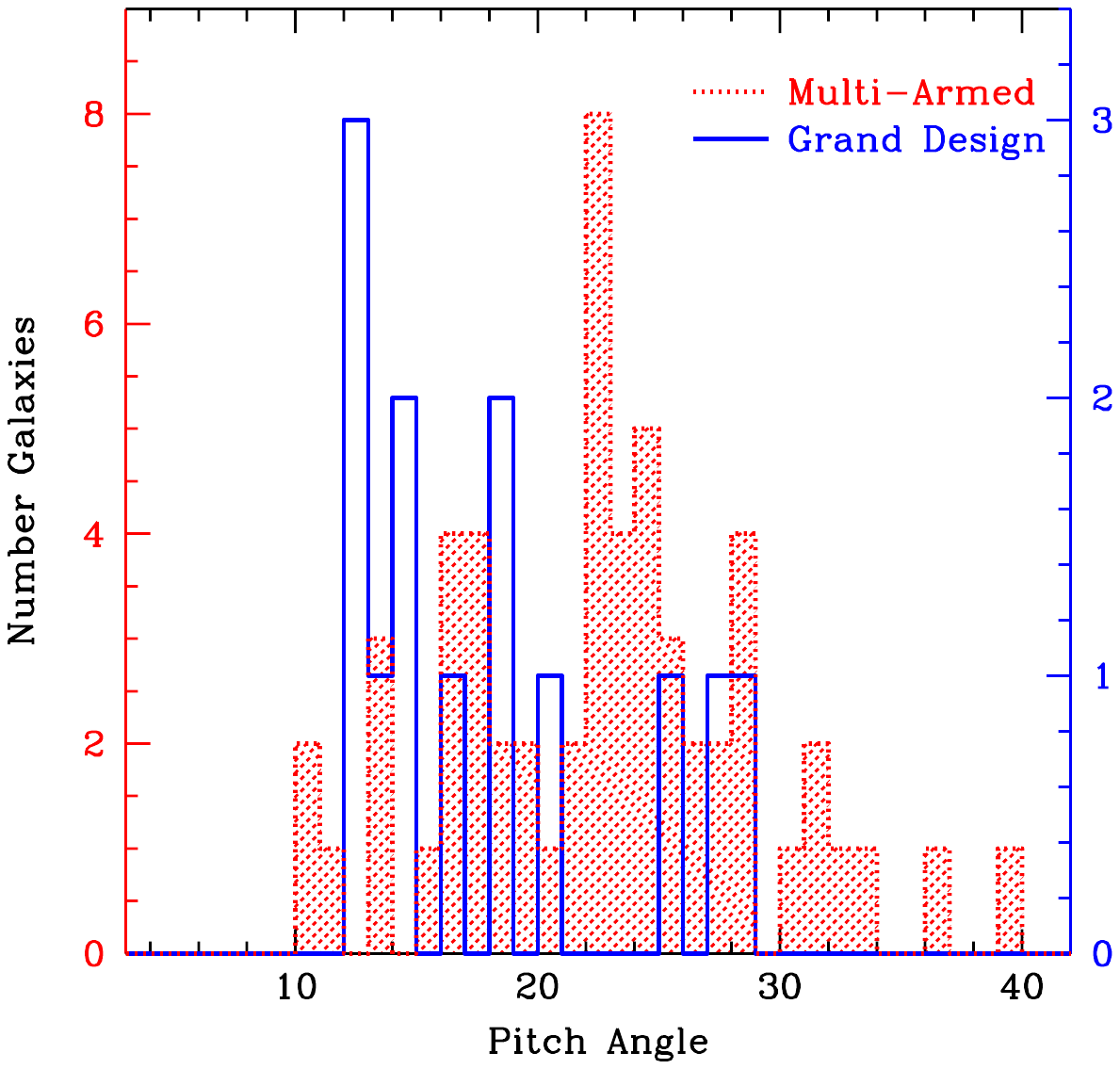}{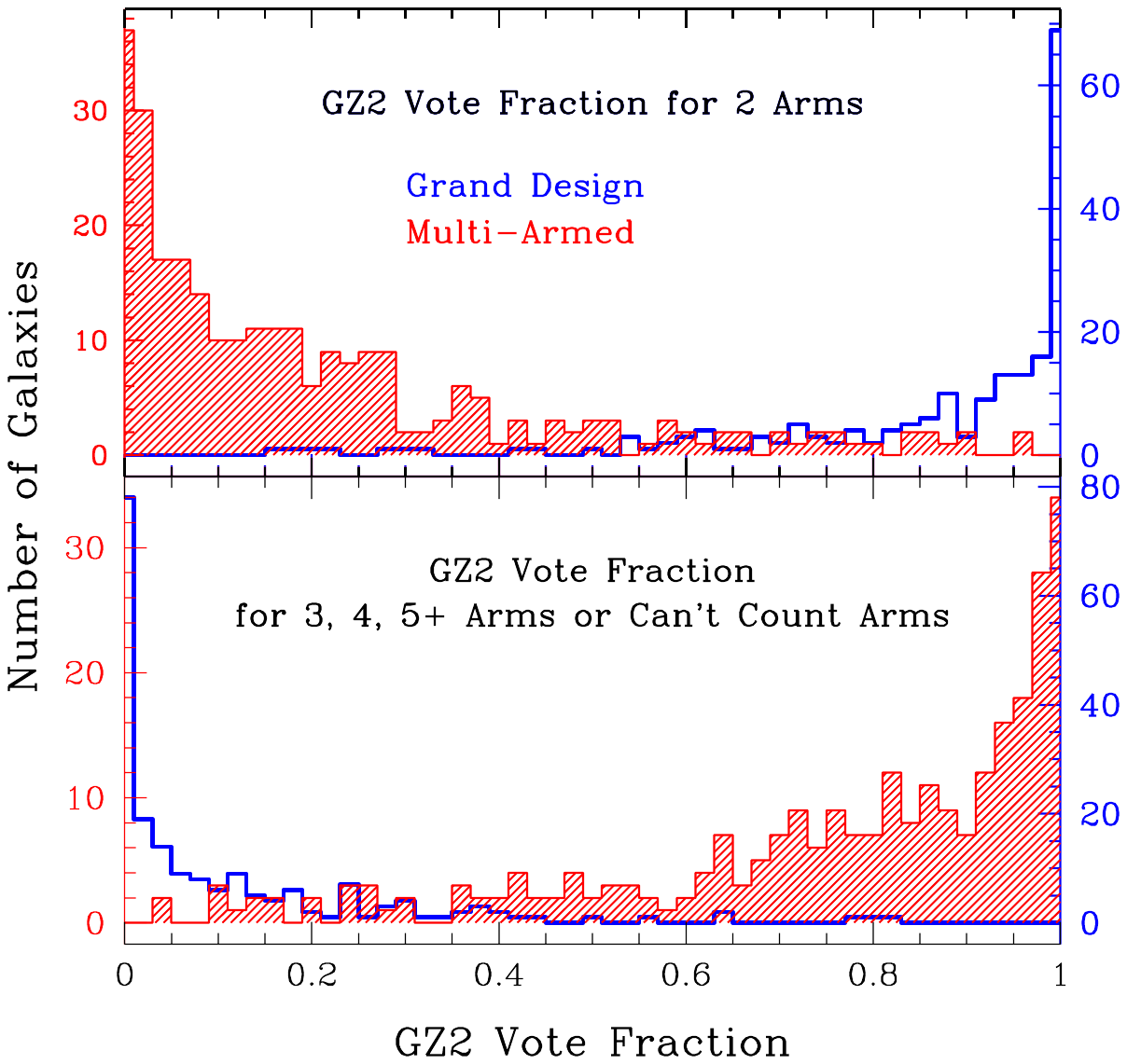}
\caption{ 
Left: Histograms of the \citet{2020ApJ...900..150Y} pitch angle 
for galaxies in our multi-armed galaxies (red hatched)
and our grand design galaxies (blue).
Right: Histograms of the GZ2 vote fraction
for 2 arms (top panel) and for either 3 arms, 4 arms, 5+ arms, or `can't count
arms' (bottom panel)
for 
galaxies in our grand design sample (blue)
and our multi-armed sample (red hatched).
Only `secure spirals' according to the 
definition of 
\citet{2013MNRAS.435.2835W}
which have
at least 20 classifications in GZ2 are 
included in the histograms on the right.
\label{fig:pitchangle_GZ2}}
\end{figure}

In the right panel of Figure \ref{fig:f3}, we provide histograms
of the \citet{2020ApJ...900..150Y} arm strengths for the 
multi-armed 
vs.\ grand design galaxies. 
The KS probability is 0.074, thus we cannot rule out that the two
distributions come from the same parent sample.
Histograms of the \citet{2020ApJ...900..150Y} 
pitch angles for our multi-armed vs.\ grand design
galaxies are given in 
the left panel of 
Figure \ref{fig:pitchangle_GZ2}.  
While the range of pitch angle is large for each class,
the median pitch angle is lower
for the grand design galaxies.  
This implies that they tend to be more tightly wound than the multi-arm galaxies.
This trend may be a consequence of the relation between
concentration and number of arms (Figure \ref{fig:NSAconcentration}), 
and an underlying correlation between
pitch angle and concentration \citep{2020ApJ...900..150Y}.
At any rate, the difference is marginally significant
(KS probability 0.035).

\subsection{Comparison with Galaxy Zoo 2 Arm Count Vote Fractions } \label{sec:GZ2}

We also compared our arm counts with the results of the Galaxy Zoo 2
(GZ2)
citizen scientist project
\citep{2013MNRAS.435.2835W}.
Galaxy Zoo 2 asked participants a series of questions about each galaxy.
First, participants were asked whether the 
galaxy is `smooth', has `features/disk', or has an `artifact'. 
If `features/disk' was chosen, 
then 
they were asked if it is an edge-on disk.
If `not edge-on' was selected, then participants were asked
whether the galaxy has spiral arms.  If the participant said spiral arms
were present, 
they were then 
asked how many spiral arms are present, with the choices being
1, 2, 3, 4, or 5+, or `can't count arms'.

\citet{2016MNRAS.461.3663H} define a `secure spiral' as a galaxy
for which 
p$_{\rm feature}$$\times$p$_{\rm not~edge-on}$$\times$p$_{\rm spiral}$ $>$ 0.5,
where p$_{\rm feature}$ is the fraction of votes in GZ2 for features/disk,
p$_{\rm not~edge-on}$ is the fraction of votes for not being edge-on,
and p$_{\rm spiral}$ the fraction of votes for having spiral arms.
Of our set of 299 multi-armed and 245 grand design galaxies, 
272 and 194, respectively, meet this condition and have at 
least 20 classifications
votes in GZ2.

In the top right panel of Figure \ref{fig:pitchangle_GZ2},
for our multi-armed and grand design galaxies that meet these
criteria,
we display histograms of the fraction of the
GZ2 volunteers who selected `2 arms' in GZ2, while 
the bottom right panel 
shows the fraction who selected
either 
`3 arms', `4 arms', `5+ arms', or `can't count
arms'.   This comparison shows that the vast majority of GZ2 viewers
agreed that the grand design galaxies have two arms, while
the majority of the GZ2 voters
selected either 3 arms, 4 arms, 5+ arms, or `can't
count arms' for our multi-armed galaxies.
This shows good consistency between our method of counting spiral
arms and the GZ2 statistics.

\section{Discussion} \label{sec:discussion}

\subsection{Comparison with Earlier Studies}

We developed an algorithm for using 
GZ-3D masks to count the number of spiral arms
as a function of radius within galaxies, and created
a catalog of 
299 multi-arm galaxies and 245 grand design galaxies.
For galaxies that overlap with other studies,
our classifications are in reasonable agreement with previous
classifications.
We compared the concentrations, stellar masses, Hubble types,
central surface mass densities, and sSFRs of our two
classes of spirals, and reached a number of conclusions. 
In this section, we compare our main conclusions to those of past studies.

In this study, we focus on grand design vs.\ multi-arm galaxies,
and do not specifically target flocculent galaxies or treat them
as a separate class.   
Our initial target list, the GZ-3D sample, 
explicitly excludes galaxies 
identified by GZ2 viewers as having more than four 
arms, thus our initial selection criteria 
was biased against flocculent galaxies. 
Furthermore, our requirement that the GZ-3D masks have 
continuous radial sequences 
with the same number
of arms also biases the search against fragmented arms. 
Some galaxies
traditionally classified as flocculent may have 
ended up in our final multi-armed sample, although based
on our by-eye inspection of the images the percentage
of true flocculent galaxies in our multi-armed sample is relatively small.
In our visual inspections of 
the images, we noted that many galaxies in both the grand design and the
multi-armed sample appear clumpy in the SDSS g images, probably because
of prominent knots of star formation along the spiral arms.  
It is possible that some of the
galaxies we classified as multi-armed
might be judged to be flocculent by other researchers, since
the distinction between multi-armed and flocculent morphology
is somewhat subjective.
This point should be kept in mind 
in the following discussion, where we 
compare our results with the results of earlier studies in which 
arm classes were determined by eye. 

\subsubsection{Arm Counts vs.\ Concentration and Hubble Type}

We see clear differences in the 
concentrations
of the grand design and the multi-armed
galaxies in our sample.
For a given stellar mass, grand design galaxies tend to 
have larger concentrations
compared to multi-armed galaxies.
This difference in C becomes particularly pronounced 
at stellar masses greater than 10$^{10}$~M$_{\sun}$.
Compared to multi-armed galaxies, there is a deficiency of 
high mass, low concentration grand design galaxies in our sample.
For both the multi-armed and the grand design galaxies,
we see a trend towards earlier Hubble types for larger 
stellar masses.

Our conclusion that 
grand design galaxies have larger concentrations
and earlier Hubble types than multi-armed galaxies
agrees with most earlier studies
on the topic
\citep{1982MNRAS.201.1021E, 2017MNRAS.471.1070B, 2020ApJ...900..150Y}
but not all 
\citep{2013JKAS...46..141A}.
In the original formulation of the arm class system
as presented by
\citet{1982MNRAS.201.1021E},
spiral galaxies were divided into 12 classes based on 
arm
morphology.   
Classes 1 $-$ 4 are varieties
of 
flocculent galaxies, while classes 10 $-$ 12 are
classical 2-armed grand design galaxies.
Galaxies in classes 5 $-$ 9 include systems similar to those
we label multi-armed galaxies in this study.
For their sample of 305 spiral galaxies,
\citet{1982MNRAS.201.1021E} plot Hubble type vs.\ their
12 arm classes in their Figure 1.
The flocculent galaxies in classes 1 $-$ 4 tend to be
very late types, while class 12, the most common of the grand
design galaxies, tend to have earlier types.
Arm classes 5 and 9, the most common types in their class 5 $-$ 9 
range, have Hubble types between those of the flocculent galaxies
and the grand design galaxies.
This is consistent with our results on the
Hubble types of grand design vs.\ multi-armed galaxies.
Many of our multi-armed galaxies have 
morphologies consistent with the 
\citet{1982MNRAS.201.1021E} 
arm class 9 definition 
(`multiple long arms in the outer parts, two symmetric
and continuous arms in the inner parts') or arm class 5 definition
(`two symmetric short arms in the inner regions; irregular
arms in the outer regions'). 

For a sample 
of 1064 nearby spiral galaxies 
classified
by eye, 
\citet{2017MNRAS.471.1070B} found statistically-significant
differences in the TTypes of their grand design, multi-armed, 
and flocculent galaxies.
Their
grand design galaxies have the lowest TTypes,
followed by multi-armed galaxies, with flocculent galaxies
having the highest TTypes.  
Excluding their flocculent galaxies,
our results are 
consistent with theirs.

\citet{2013JKAS...46..141A}
visually classified
a set of 1725 SDSS spiral galaxies with z $<$ 0.02
into grand design, flocculent, and multi-armed galaxies.
In contrast to our results and the results of 
\citet{2015ApJS..217...32B},
their grand design galaxies have similar or
slightly smaller concentrations than their
multi-armed galaxies.
\citet{2013JKAS...46..141A}
also found that the distribution of
Hubble types for their multi-armed galaxies tended 
to be peaked at earlier types than the distribution
for grand design galaxies, also inconsistent with our
results.  
This difference may be due to different ranges
in M*.
Their 
SDSS sample is limited to angular sizes greater than 10$''$,
and otherwise is complete to about 
an absolute magnitude in r of 
M$_{\rm r}$ = $-$16.1.  
To convert this absolute magnitude into an approximate stellar
mass, 
we derived a best-fit linear relation between log M*
and M$_{\rm r}$ for all field galaxies in the NSA
of log (M*/M$_{\sun}$) = $-$0.49 M$_{\rm r}$ + 0.03. 
We defined field galaxies 
as galaxies that
are not within 10 Mpc of
the center of any
\citet{2017MNRAS.470.2982L}
group or cluster
with a halo mass greater than 10$^{12.5}$~M$_{\sun}$ 
and more than four members.
From this relation, we estimate that 
the 
\citet{2013JKAS...46..141A} absolute magnitude limit
corresponds to about log (M*/M$_{\sun}$) = 7.9. 
Their sample thus extends to lower stellar masses than our 
sample. 
Since both TType 
and concentration
depend upon stellar 
mass 
(Figures \ref{fig:C_vs_mass}
and \ref{fig:DLHubType}),
their sample may be skewed to later types and smaller concentrations
than our sample.
At the low mass end, there is less difference in C between
multi-armed and grand design galaxies
(Figure \ref{fig:C_vs_mass}).
The inclusion of very low mass galaxies in the 
\citet{2013JKAS...46..141A} sample may explain why
they found different concentration and TType distributions
relative to our sample.

Our results on concentration agree with those of \citet{2020ApJ...900..150Y},
who found a weak anti-correlation between the normalized
m=3 Fourier amplitude f3 and concentration.  This implies
that galaxies with larger bulges are more likely to have two symmetric
arms.

\subsubsection{Arm Counts vs.\ Stellar Mass}

A second major conclusion of our paper is that grand design
galaxies tend to have smaller stellar masses than multi-armed galaxies.
This result agrees with earlier GZ2 studies.
Using GZ2 arm counts for galaxies with 
9 $\le$ log (M*/M$_{\sun}$) $<$ 11, 
\citet{2022MNRAS.515.3875P} found that multi-armed galaxies
with 3, 4, or more arms tend to have higher stellar masses
than 2-armed galaxies.   In an earlier GZ2 study
focused on high mass galaxies (log (M*/M$_{\sun}$) $\ge$ 10.6), 
\citet{2016MNRAS.461.3663H} found that the fraction of spirals
that are multi-armed increased with M* relative to 2-armed galaxies.
We find a peak in the ratio of the number of multi-armed galaxies to 
the number of grand design galaxies
at log (M*/M$_{\sun}$) = 10.3 $-$ 10.6 
(see Appendix \ref{sec:selection}), 
although
the ratio drops off at higher M*.

Our result on stellar masses differs from that 
of
\citet{2017MNRAS.471.1070B}, however.
For their sample of
1064 galaxies classified by eye into the three basic arm classes,
\citet{2017MNRAS.471.1070B} found 
no significant difference
between
the stellar masses of their multi-armed and grand design galaxies.
The difference may be sample selection.  
Their sample 
is a subset of a volume-limited, magnitude-limited,
and angular-size-limited sample
(distance $<$ 40 Mpc, B magnitude $<$ 15.5, D$_{25}$ $>$ 1$'$).
Most of their galaxies lie between 9 $\le$ log (M*/M$_{\sun}$) $<$ 10, 
with only a relatively small fraction above log (M*/M$_{\sun}$) = 10. 
Our galaxies are more distant, with larger masses on average.

\subsubsection{Size vs.\ Stellar Mass Relations}

\begin{figure}[ht!]
\plotone{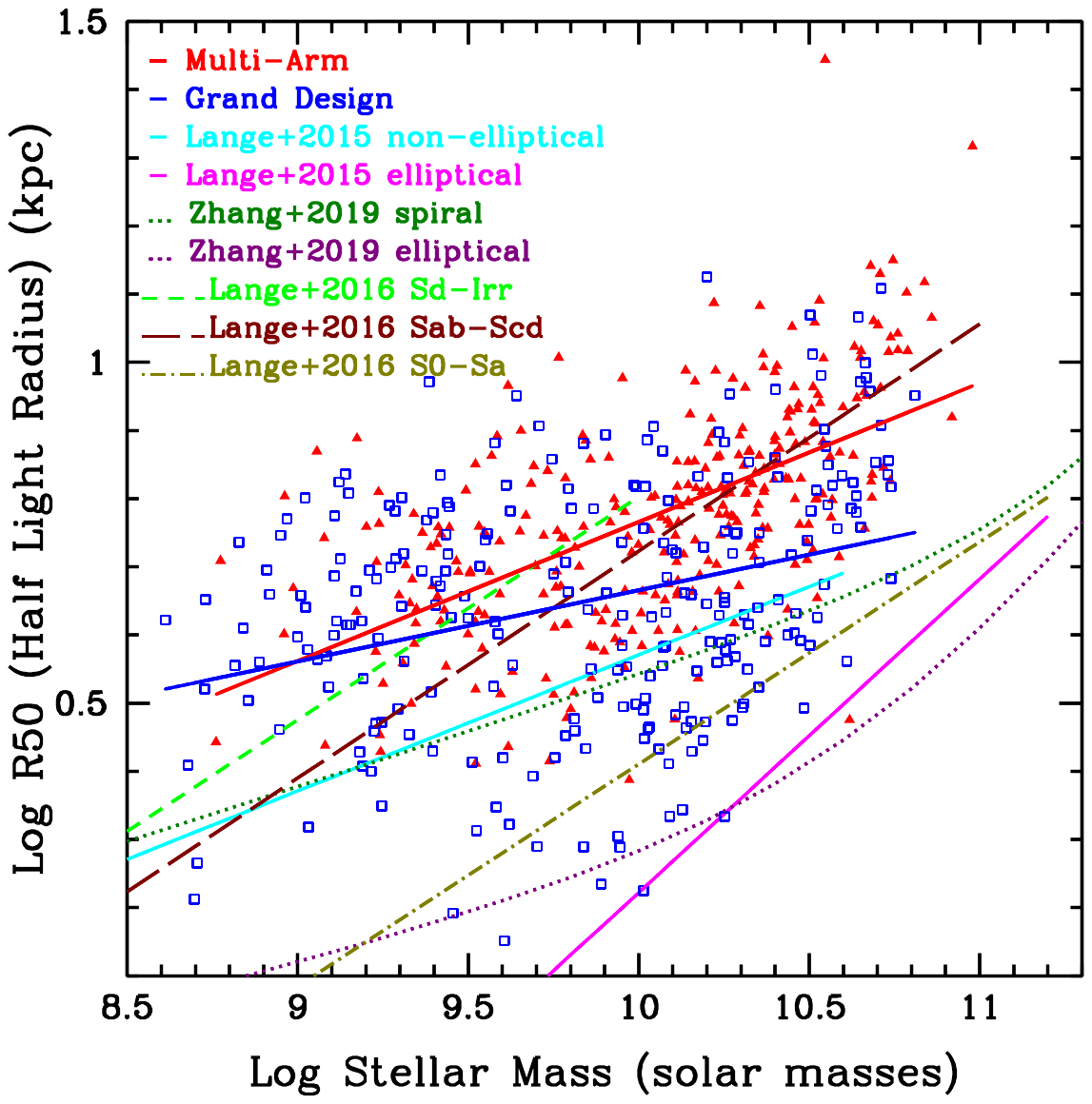}
\caption{ 
Plot of R50 vs.\ stellar mass for our multi-armed (red) vs.\ 
grand design galaxies (blue), as in 
Figure \ref{fig:Size1}.  Results from other studies
are 
overlaid.   See text for details.
\label{fig:Reff_overlay}}
\end{figure}

Our grand design galaxies have a flatter 
log R50
vs.\ log M* relation than the multi-armed galaxies.
At higher masses, the grand design galaxies have smaller
R50 for a given M* than multi-armed galaxies.
In Figure \ref{fig:Reff_overlay}, 
we compare our datapoints and our best-fit relations
to those of earlier r-band studies.
For a sample of 8399 galaxies in the GAMA survey with 0.01 $<$ z $<$ 0.1,
\citet{2015MNRAS.447.2603L}
fit log R50 vs.\ log M* for ellipticals and non-ellipticals.
The slope of their best-fit relation for the non-ellipticals
(0.20 $\pm$ 0.02)
is the same as ours for the multi-armed galaxies,
however, their line is offset to lower radii.
Their elliptical galaxies show a steeper slope, but for
log (M*/M$_{\sun}$) $<$ 11 the ellipticals have significantly
smaller effective radii than non-ellipticals for a given 
stellar mass.
In Figure \ref{fig:Reff_overlay}, we also display the best-fit
relations for a sample of SDSS ellipticals and spirals from  
\citet{2019RAA....19....6Z}.
Their relations agree well with those of 
\citet{2015MNRAS.447.2603L}.

In Figure \ref{fig:Reff_overlay}, we also overlay
the \citet{2016MNRAS.462.1470L}
best-fit lines for 
S0-Sa, Sab-Scd, and Sd-Irr galaxies
for a sample of 
GAMA galaxies with 0.002 $<$ z $<$ 0.06.
All three morphological classes have similarly steep
slopes of about 0.33, 
however, the 
y-intercept of the relation
increases from early to late type (i.e., later type galaxies have larger
effective radii for a given mass than earlier types).
Our relation for multi-arm galaxies is flatter than their relations,
but agrees well at the high mass end with that of Sab-Scd galaxies.
Our grand design galaxies mostly lie between their relations for Sab-Scd 
and 
S0-Sa galaxies at the high mass end, but above their relation
for Sab-Scd at the low mass end. 

Our samples appear deficient
in low mass, small size galaxies (log R50 $<$ 0.04 and 
log (M*/M$_{\sun}$) $<$ 9.5) 
compared to the \citet{2016MNRAS.462.1470L} best-fit lines.
This deficiency is not a consequence of how the half light radius
was determined;
using the NSA S\'ersic half-light radius rather than
the Petrosian radius shifts our relations to higher radii,
making the offset even larger. 
One factor that may contribute to this deficiency is
our selection criteria.
For grand design galaxies 
we are limited to 
radii $\ge$ 7$''$, due to our requirement that the spiral arms must
be traced out at least 5$''$ in the galaxy, starting at a radius of 2$''$.
Assuming 
a relatively low concentration of C = 2, and assuming that the arms
extend out to R90, this means that R50 must be $\ge$3\farcs5. At
a distance of z = 0.02, this corresponds to R50 = 1.5 kpc, or
log R50 = 0.18.   Thus the bottom area on these plots is not accessible
in our study for low redshift, low concentration galaxies. 
However, higher concentration galaxies could fall in this region.

However, the main reason for the flatter slopes for our samples 
in Figure \ref{fig:Reff_overlay}
compared to
the relations for S0-Sa, Sab-Scd, and Sc-Irr galaxies
is 
the anti-correlation between TType and stellar
mass 
(Figure \ref{fig:DLHubType}).
For both grand design and multi-armed galaxies,
lower mass galaxies have later TTypes and smaller concentrations.
Our low mass 
galaxies of both arm classes are close to the
\citet{2016MNRAS.462.1470L} relation for late-type spirals because they 
have small concentrations, and preferentially are classified as
late-type spirals. 
Both our grand design sample and our multi-armed sample 
lack low mass galaxies classified as type S0-Sb 
(Figure \ref{fig:DLHubType}).

\subsubsection{Pseudo-Bulges vs.\ Classical Bulges} \label{sec:pb_cb}

Based on the criterion log $\Delta$$\Sigma$$_1$ $>$ 0
from \citet{2020MNRAS.493.1686L},
70\% of our grand design galaxies host classical bulges, but only 50\% of 
the multi-armed galaxies.
The question of pseudo-bulges vs.\ classical bulges in grand design,
multi-armed, and flocculent galaxies was 
addressed before by 
\citet{2017MNRAS.471.1070B}, 
by fitting 
S\'ersic profiles to the bulges of their spiral galaxies.
According to 
\citet{2004ARA&A..42..603K} and \citet{2008AJ....136..773F},
pseudo-bulges and classical bulges can be distinguished by 
the 
bulge S\'ersic index n; n $<$ 2
signals a pseudo-bulge, while n $>$ 2 suggests 
a classical bulge.
\citet{2017MNRAS.471.1070B} 
found that the 
distributions of bulge S\'ersic indices 
for grand design and multi-armed galaxies extend
to high n,
however, less than 
half of their galaxies have n $>$ 2.   
They found similar distributions of bulge S\'ersic indices for the
grand design and multi-armed galaxies, implying similar numbers
of classical vs.\ pseudo-bulges.

As noted earlier, our sample is skewed to higher stellar masses
than the 
\citet{2017MNRAS.471.1070B} sample.  
We might therefore expect to see 
higher fractions of classical bulges than
the \citet{2017MNRAS.471.1070B} study.
The percentage of classical
bulges is known to increase with increasing stellar mass.
Classical bulges dominate
above log (M*/M$_{\sun}$) $\sim$ 10.5, 
while pseudo-bulges are more common at lower
masses
\citep{2020ApJ...899...89S,
2023arXiv231202721H}.
The tendency of grand design galaxies to have earlier Hubble
types and larger concentrations for a given mass
may also favor 
classical bulges.
Classical bulges are more frequent in early-type spirals,
although they 
are seen in the full range of spiral Hubble types
\citep{2004ARA&A..42..603K, 
2007MNRAS.381..401L}.
Concentration is also 
weakly correlated with bulge S\'ersic
index,
although concentration alone cannot
be used to distinguish between pseudo-bulges
and classical bulges
\citep{2009MNRAS.393.1531G}.

As an independent check on the nature of the bulges
in the centers of our galaxies, we cross-correlated
our list of galaxies with the 
\citet{2022MNRAS.509.4024D}
catalog of photometric parameters for MaNGA galaxies.
Their catalog includes the results of fits to the SDSS
surface brightness profiles of the galaxies.
Following \citet{2023arXiv231202721H},
for this comparison we limited 
the sample to galaxies 
with a reliable 2-component
fit (bulge plus disk)
by setting the parameters FLAG$\_$FIT=2
and FLAG$\_$FAILED$\_$SE=0, and requiring the
half-light radius for the disk component to be larger than that for the bulge
component and the half-light radius of the bulge to be
larger than 0\farcs75 (about the half width half maximum point
spread function).  
A total of 41 and 32 of our multi-armed and grand design galaxies,
respectively, meet these criteria.
Of these, eight 
(20\%) of the multi-armed and nine (28\%) of the grand
design have a bulge S\'ersic index 
larger than 2.  
These percentages are smaller than those implied by the 
$\Delta$$\Sigma$$_1$ analysis, however, the sample sizes are small,
and the uncertainties on the bulge
S\'ersic indices are sometimes quite large.
A more detailed study of the radial light profiles in these
galaxies using higher spatial resolution images would be helpful
to further test the idea that grand design galaxies are more likely
to host classical bulges.

\subsubsection{sSFR vs.\ Arm Counts }

We find no relation between arm count and sSFR, if one removes the
effect of concentration (i.e., comparing galaxies with similar
concentrations and stellar masses, we see no difference between
the sSFRs of grand design and multi-armed galaxies.)
This conclusion agrees with that
of \citet{2022AJ....164..146S}, 
who found that the m=3 Fourier amplitude
of spiral galaxies
is weakly correlated with sSFR, but when accounting for
differences in concentration and stellar mass, no trend is present.
We also see little difference in the sSFRs of multi-armed and grand design
galaxies if we compare galaxies of similar stellar masses and 
similar central surface brightnesses.

There have been a number of GZ2 studies investigating how
sSFR depends upon the number of spiral arms.
Some studies concluded that 3-armed systems have higher
sSFRs than 2-armed galaxies
\citep{2016MNRAS.461.3663H, 
2022MNRAS.517.4575S},
other studies
\citep{2015MNRAS.449..820W,
2017MNRAS.468.1850H}
did not see a statistical difference
in sSFR between galaxies with 2 arms and galaxies with 3 arms,
and one study
\citep{2022MNRAS.515.3875P}
concluded that multi-armed galaxies have lower sSFR.
These different results might
be a consequence of sample effects,
since each study includes galaxies with different M* ranges,
and sSFR is a function of both M* and concentration.
The 
\citet{2016MNRAS.461.3663H}
and 
\citet{2022MNRAS.517.4575S} 
samples are limited to
high mass galaxies (log (M*/M$_{\sun}$)  $\ge$ 10.6 
and 10.25 $\le$ log (M*/M$_{\sun}$) $<$ 10.75, 
respectively). Within these mass ranges, when galaxies with
all concentrations are included we find 
sSFRs that are 
significantly higher
or marginally significantly 
higher 
for multi-armed galaxies than for grand design galaxies,
consistent with their results.
However, in this mass range there 
are significant differences between the concentrations
of grand design and multi-armed galaxies 
(Table 
\ref{Ctab})
and sSFR also depends upon concentration.  When we remove the effect
of concentration by comparing only galaxies with similar concentrations,
the sSFRs of multi-armed galaxies are similar to those of grand design
galaxies 
(Figure 
\ref{fig:mainseq}).
We conclude that the differences in sSFR
between 3-armed and 2-armed galaxies
found by 
\citet{2016MNRAS.461.3663H}
and 
\citet{2022MNRAS.517.4575S} 
are a consequence of differences in concentration between the
two sets of galaxies.

The \citet{2022MNRAS.515.3875P} GZ2 study includes galaxies
with a wide range of stellar mass,
so here we need to account for
the effect of averaging over a range of masses.
Their sample included galaxies with 
9 $\le$ log (M*/M$_{\sun}$ $<$ 11, and was strongly biased
towards lower mass galaxies.
Since their 2-armed galaxies have lower masses on average
than their 3+armed galaxies (in agreement with our results)
and sSFR tends to decrease with increasing M* (Figure 
\ref{fig:mainseq}),
weighting over their mass distribution 
may cause
their 
2-armed galaxies to have 
higher sSFRs than 
the
3+armed galaxies on average.

In the \citet{2015MNRAS.449..820W} 
and \citet{2017MNRAS.468.1850H} studies,
the sSFR variation with M* is investigated
explicitly for 2-armed galaxies vs.\ 3+armed galaxies.
They reach the same conclusion as we do, that the trend
with M* is independent of the number of arms.

\subsection{Comparison with Theories of Spiral Arm Production and Maintenance} \label{sec:models}

Statistical studies of galaxies with multi-armed vs.\ grand design
patterns like the current study
provides important constraints
on theories of spiral arm formation and maintenance. 
In this section, we discuss the various models
of spiral arm development in light of the data.

The sample of galaxies studied in the current paper contains very few strongly
interacting pairs or cluster galaxies, so we cannot directly
compare with earlier studies of interacting or cluster galaxies.
A more detailed study of
the environment around each galaxy in the current sample is beyond the
scope of this study, and we leave that to future work.
Ideally, since arm class is a function of both concentration
and stellar mass, this study would be repeated with a large
enough sample to allow separation into subsets based on M* and C
as well as environment.
In any case, our observation that 
grand design galaxies are systematically offset 
from multi-armed galaxies 
in the concentration-M* plane
strongly suggests that the specific spiral pattern observed
in a galaxy depends at least in part on the internal structure of the 
galaxy, and not just on external influences.
Although gravitational interactions are clearly responsible
for some spiral patterns, they are likely not the sole explanation
for the difference between grand design and multi-armed galaxies.

We have not included an analysis of bars 
in the current sample of galaxies in this paper;
that task is also left to future work.
Past studies on the statistics of bars in spirals
give ambiguous results.
Although correlations are seen between bar strength and arm strength 
in spirals
\citep{2004AJ....128..183B, 2005AJ....130..506B, 2020ApJ...900..150Y},
it is unclear whether the bar causes the arms, or whether conditions
that favor bar production also favor strong arms
\citep{2010ApJ...715L..56S, 2019A&A...631A..94D}. 
In any case, the presence of grand design spiral arms in unbarred
isolated galaxies implies that another process is necessary for at least in
some galaxies.

Observationally, we find that 
grand design galaxies
are more likely to
host classical bulges than multi-armed galaxies,
and have larger concentrations for a given stellar mass.
This suggests that the presence of a dense concentrated bulge
aids the development and/or 
longevity of a 2-armed spiral pattern.
This is consistent with the idea that a massive bulge may reflect a
spiral wave and make it long-lived, as suggested
by   
\citet{1989ApJ...338..104B}.
In this scenario, a trailing spiral wave moving inwards reflects off
of the high Q barrier caused by the bulge, and returns outwards as 
a weak leading wave 
to corotation, at which point swing amplification reverses it again
\citep{1976ApJ...205..363M, 1977ApJ...212..645M, 
1989ApJ...338..104B}.
Under the right conditions, a long-lived spiral mode can be produced.
The N-body simulations of 
\citet{2016ApJ...826L..21S}
appear to
capture this process, 
as they produce a strong 2-armed spiral mode wave with a constant pattern
speed that lasts 4 $-$ 5 Gyrs.

The formation and evolution of large scale spirals
in galaxies 
can be modeled in terms of ensembles of resonant eccentric
orbits 
(\citealp{2015MNRAS.450.2217S, 
2024MNRAS.528.7492S}, in prep).
These orbit calculations show that it is easier to get 2-arm-supporting
resonant eccentric orbits 
over a range of pattern speeds and radii
when the rotation curve is falling or flat,
as opposed to rising.
In contrast, resonance eccentric orbits associated
with m=3 and m=4 spirals are favored over a range
of radii when the rotation curve is rising.
This suggests that a centrally concentrated 
gravitational potential (associated with a flat or falling rotation
curve) favors 2-armed spiral patterns, and a more extended
mass distribution favors multi-armed structure.
This is consistent with our observational results.

In contrast to the 
long-lived spiral waves in the 
\citet{2016ApJ...826L..21S}
simulations,
simulations 
in which spirals are caused
by instabilities in the disk enhanced by swing amplification
produce transient and recurrent arms 
\citep{1984ApJ...282...61S,
2000Ap&SS.272...31S,
2009ApJ...706..471B,
2013ApJ...763...46B,
2013ApJ...766...34D,
2018MNRAS.481..185M}.
Although these simulations frequently produce multi-armed
and flocculent structures, 
two-armed morphologies can also appear, 
especially 
when the
disk is massive compared to the halo 
\citep{2015ApJ...808L...8D, 2018MNRAS.481..185M}
and/or the 
shear rate is high 
\citep{2018MNRAS.481..185M}.
A falling rotation curve implies a high shear rate,
thus these simulations show quantitative agreement with the 
orbit calculations of \citet{2024MNRAS.528.7492S}.

With simulations
it is difficult to produce an unbarred 2-armed galaxy,
since 
a massive disk tends to be unstable to bar formation
\citep{1982MNRAS.199.1069E,
2003MNRAS.344..358B, 2018MNRAS.477.1451F}, and 
a massive halo
tends to suppress both bar formation and the formation
of a 2-armed spiral \citep{2019MNRAS.486.4710S, 2023ApJ...958..182S}.
A possible way to prevent bar formation is by adding a classical bulge
\citep{1985MNRAS.217..127S}.
Orbit calculations suggest that a centrally concentrated gravitational
potential may not support very eccentric resonance orbits
in their inner regions
\citep{2024MNRAS.528.7492S},
suggesting, in turn, that it is hard to form
bars in such systems.
In a recent set of simulations,
\citet{2018ApJ...858...24S} were able to produce unbarred
spiral patterns by
including a compact and dense spherical classical bulge.
This 
produces a strong
inner Lindblad resonance
(ILR) with a small radius, which inhibits the growth of a bar.
The large ratio of random motions to ordered motion in the 
central region of the galaxy may stabilize the inner disk and prevent
bar formation.   
More recently,
\citet{2023ApJ...942..106J} ran a series of simulations with
a range of bulge/disk mass ratios and bulge and halo sizes.
Some of these simulations produced spirals
without bars; the barless models tend to have large masses
within the inner 0.1 kpc relative to the disk mass.
Most of 
the \citet{2023ApJ...942..106J} 
simulations of unbarred spiral galaxies
produce multi-armed spiral patterns,
but a subset appear to have two dominant arms.  
The
models with the two dominant arms 
have the largest
bulge/disk mass ratios.

These models suggest that the number of arms in a galaxy
is a function of the relative masses and sizes of the 
bulge, the disk, and the halo.  
Our observations show that the concentration and bulge properties
of a galaxy play important roles in determining the number 
of arms present.
Although we do not have direct evidence of 
long-lived wave modes operating in
our sample of galaxies, the fact that grand design patterns are 
more common in galaxies with large bulges is suggestive,
and supports the hypothesis that a large bulge favors
the production of two-armed spirals, or increases their longevity.

More work 
on the topic of arm count vs.\ galaxy structure is needed
on both the theoretical
and the observational side.
Large statistical studies of the rotation curves
of spirals with different arm morphologies would better
determine how the arm morphology relates to halo, disk, and bulge
mass distributions in galaxies.
Future studies of arm morphology as a function of stellar mass,
concentration, Hubble Type, and rotation curve may provide clues 
to the merger and star formation histories of galaxies.
Combining
the known relation between
stellar mass and 
the ratio of stellar mass to halo
mass 
\citep{2013ApJ...770...57B,
2006MNRAS.368..715M}
with observed trends in 
the grand design vs.\ multi-arm
fraction with stellar mass, concentration, and TType
may also help constrain
galaxy evolution models.

Above stellar masses of $\sim$10$^{10}$~M$_{\sun}$, a marked
increase in the number of spirals with high concentrations has been observed,
producing a `bend' in the concentration-M* diagram for spirals
\citep{2020MNRAS.493.1686L, 2022AJ....164..146S}.
This bend is shown in our Figure
\ref{fig:C_vs_mass}, 
where the scatter in C increases at higher M*.
\citet{2020MNRAS.493.1686L} suggest that this
bend 
is caused by the growth of classical bulges.
Our current study shows that the high C, high M* spirals above
this bend tend to have
grand design patterns.  
Below the bend,
grand design galaxies tend to have lower observed 
concentrations
(though still larger than those of multi-armed 
galaxies of the same stellar mass).
Combined with a better theoretical understanding of 
the origins of grand design galaxies,
these observations may provide clues to how galactic bulges
evolve over time.
In the Hubble Deep Field,
few grand design galaxies are seen at z = 1
compared to the local Universe
\citep{1996AJ....112..359V}.  This result may be a consequence of
bulge growth with time.
More work is needed on how the fraction of
grand design vs.\ multi-armed galaxies varies with redshift.

\section{Summary} \label{sec:summary}

We developed a method to use Galaxy Zoo 3D spiral arm masks 
to distinguish multi-armed galaxies from grand design spirals,
and created a catalog of 299 multi-armed galaxies and 245 grand design
systems.
We found reasonable agreement between our classifications
and arm counts from GZ2
and 
published
normalized m=3 Fourier amplitudes.

Galaxies classified as grand design
are offset in the concentration-M*
plane relative to multi-armed galaxies, with larger concentrations
and smaller stellar masses on average.
Grand design galaxies also tend to have earlier Hubble
types than multi-armed galaxies, with the relation skewing
to later types at lower stellar masses.
The grand design and multi-armed galaxies have similar R90 sizes 
for a given stellar mass, but the
grand design galaxies have smaller half-light radii.
The larger concentrations seen in grand design galaxies are due
to an excess of light in the central region, 
rather than larger overall sizes.
When comparing multi-armed and grand design
galaxies with similar masses and concentrations, 
they have similar sSFRs.
Based on the central surface mass densities in these galaxies,
we conclude
that grand design galaxies are more likely to host
classical bulges, while multi-armed galaxies are more likely
to have pseudo-bulges. 

Our observations are consistent with theoretical models in which 
dense classical bulges favor the development and possibly
the longevity
of two-armed spiral patterns.  Although other mechanisms, such
as interactions and bars, may also produce grand design spirals,
the strong relation between arm class and concentration implies
that the internal structure of a galaxy is a dominant factor
in determining the number of arms.


\begin{acknowledgments}
\centerline{\bf Acknowledgments}

This research was supported by a grant from the NASA Tennessee Space Grant
consortium.
We thank the anonymous referee for valuable suggestions
that greatly improved this paper.
We also thank Si-Yue Yu and Bruce Elmegreen
for
helpful comments on this manuscript.
We greatly appreciate the work of the Galaxy Zoo-3D team and the
GZ-3D volunteers who made 
this project possible.
This publication uses data generated via the Zooniverse.org platform, development of which is funded by generous support, including a Global Impact Award from Google, and by a grant from the Alfred P. Sloan Foundation.
This research also made use of the VizieR catalogue access tool
and the cross-match service
at the Centre de Donn\'ees Astronomiques de Stasbourg,
CDS, Strasbourg, France.
This research has also made use of the NASA/IPAC Extragalactic Database (NED),
which is operated by the Jet Propulsion Laboratory, California Institute of Technology,
under contract with the National Aeronautics and Space Administration.
This research is based in part on data from the Sloan Digital Sky Survey.
Funding for the Sloan Digital Sky Survey IV has been provided by the Alfred P. Sloan Foundation, the U.S. Department of Energy Office of Science, and the Participating Institutions. SDSS-IV acknowledges
support and resources from the Center for High-Performance Computing at
the University of Utah. The SDSS web site is www.sdss.org.
SDSS-IV is managed by the Astrophysical Research Consortium for the
Participating Institutions of the SDSS Collaboration including the
Brazilian Participation Group, the Carnegie Institution for Science,
Carnegie Mellon University, the Chilean Participation Group, the French Participation Group, Harvard-Smithsonian Center for Astrophysics,
Instituto de Astrof\'isica de Canarias, The Johns Hopkins University,
Kavli Institute for the Physics and Mathematics of the Universe (IPMU) /
University of Tokyo, the Korean Participation Group, Lawrence Berkeley National Laboratory,
Leibniz Institut f\"ur Astrophysik Potsdam (AIP),
Max-Planck-Institut f\"ur Astronomie (MPIA Heidelberg),
Max-Planck-Institut f\"ur Astrophysik (MPA Garching),
Max-Planck-Institut f\"ur Extraterrestrische Physik (MPE),
National Astronomical Observatories of China, New Mexico State University,
New York University, University of Notre Dame,
Observat\'ario Nacional / MCTI, The Ohio State University,
Pennsylvania State University, Shanghai Astronomical Observatory,
United Kingdom Participation Group,
Universidad Nacional Aut\'onoma de M\'exico, University of Arizona,
University of Colorado Boulder, University of Oxford, University of Portsmouth,
University of Utah, University of Virginia, University of Washington, University of Wisconsin,
Vanderbilt University, and Yale University.

\end{acknowledgments}

\appendix

\section{Example GZ-3D Masks and SDSS Images} \label{sec:AppendixA}

\begin{figure}[ht!]
\plotone{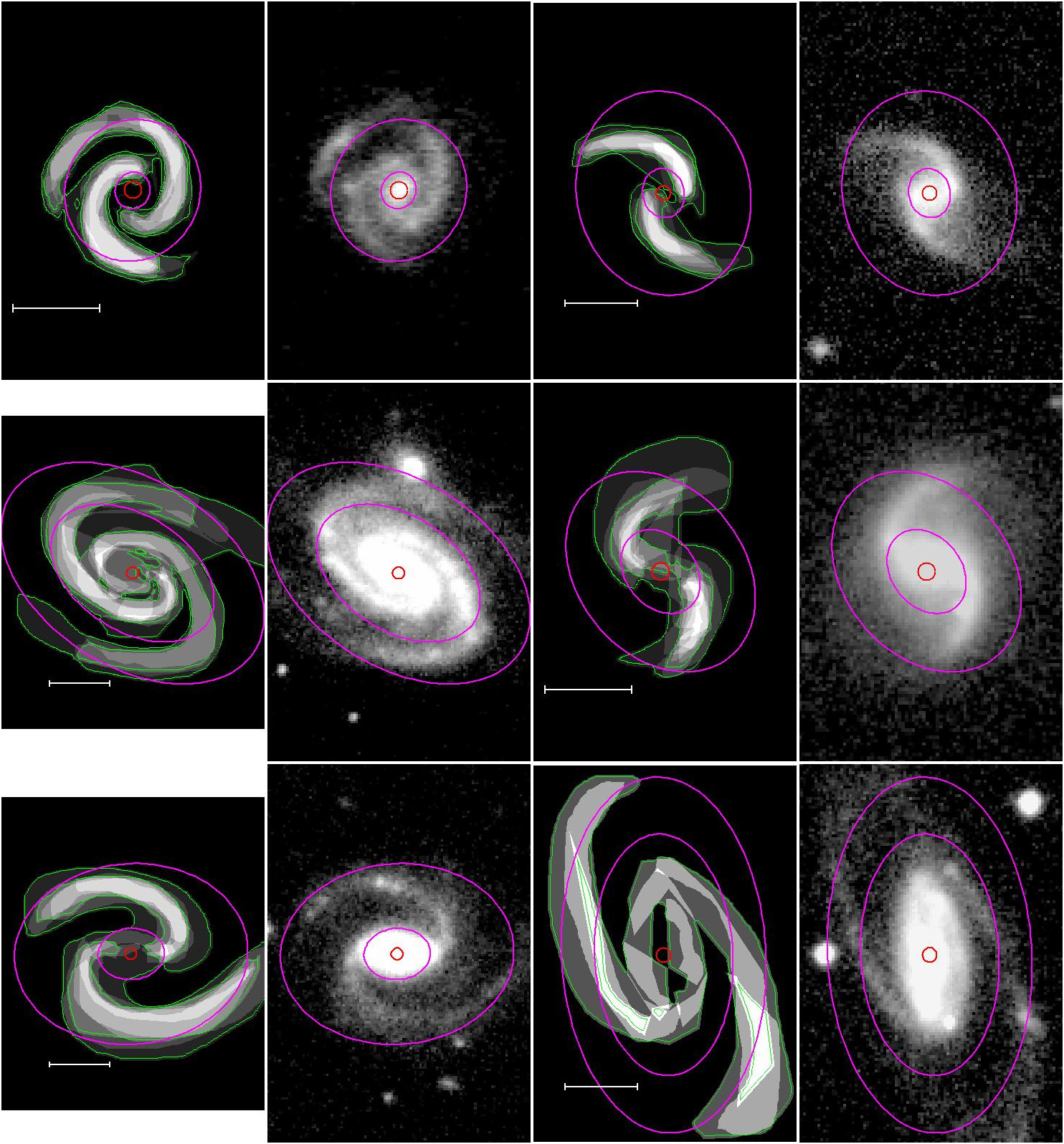}
\caption{ 
Examples of six galaxies identified by the code as grand design and confirmed
by eye.   The first picture of each galaxy is the GZ-3D mask; the second 
picture is the SDSS g-image.
The two green contours on the GZ-3D masks correspond to 1 and 3 counts.
The magenta ellipses show the inner and outer radii of the largest continuous radial
range marked as 2-armed by our software. 
The red circle indicates the center of the galaxy as marked by GZ-3D participants. 
On the SDSS g-band images, the same circle and ellipses overlaid.
In order left to right, top to bottom, the galaxies are:
MaNGA ID 1-1811 =  IAU J101927.42-000204.3;
MaNGA ID 1-32942 = IAU J010650.99-003413.7;
MaNGA ID 1-260943 = IAU J141031.43+385353.8;
MaNGA ID 1-555127 = IAU J023421.64-075224.2;
MaNGA ID 1-567320 = IAU J094120.45+211656.9;
MaNGA ID 1-605134 = IAU J081752.00+125350.9.
The white scale bars are 10$''$.
\label{fig:montage_GD}}
\end{figure}

\begin{figure}[ht!]
\plotone{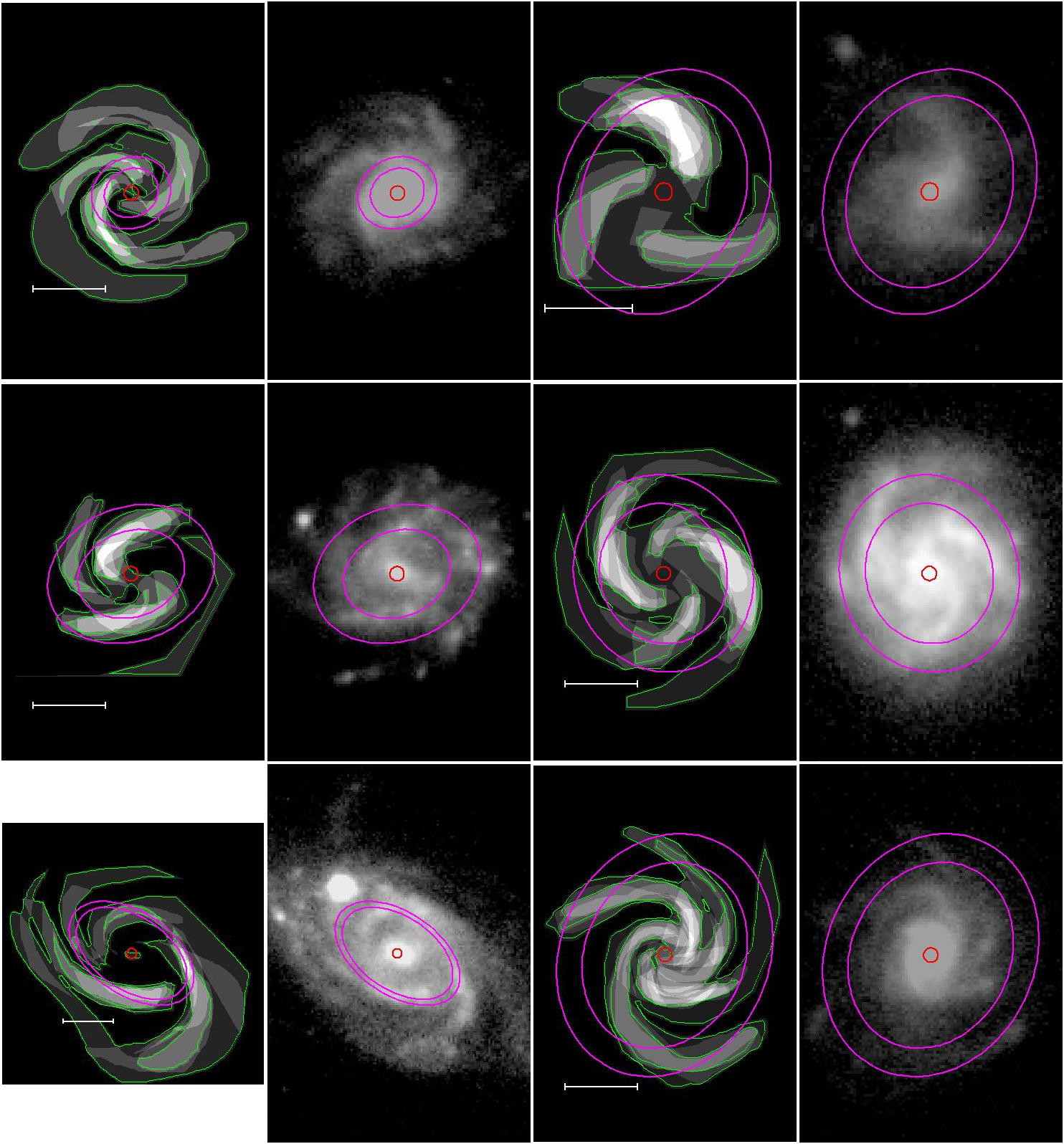}
\caption{ 
Examples of six galaxies identified by the code as multi-armed and confirmed
by eye.   The first picture of each galaxy is the GZ-3D mask; the second 
picture is the SDSS g-image.
The two green contours on the GZ-3D masks correspond to 1 and 3 counts.
The magenta ellipses show the inner and outer radii of the largest continuous radial
range marked as 3-armed by our software. 
The red circle indicates the center of the galaxy as marked by GZ-3D participants. 
On the SDSS g-band images, the same circle and ellipses overlaid.
In order left to right, top to bottom, the galaxies are:
MaNGA ID 1-30797 = IAU J001039.34-000310.3;
MaNGA ID 1-104906 = IAU J014636.04-085944.3;
MaNGA ID 1-397480 = IAU J124727.84+403359.6;
MaNGA ID 1-604111 = IAU J033758.99-061611.9;
MaNGA ID 1-638560 = IAU J121744.69+131029.9.
The white scale bars are 10$''$.
\label{fig:montage_MA}}
\end{figure}

In Figure \ref{fig:montage_GD},
we provide examples of 
galaxies in our final vetted sample of grand design galaxies.
Example galaxies from our final sample of multi-armed galaxies are shown in 
Figure \ref{fig:montage_MA}.   In these figures, we provide both the 
GZ-3D mask of the galaxy and the SDSS g image.   
On the GZ-3D masks, we have overlaid contours in green which mark the levels
of one and three counts. 
We have also overlaid ellipses
marking the inner and outer radii of the largest radial range 
in which our code counts two arms (for the grand design galaxies) or three arms
(for the multi-armed galaxies). 

\section{Possible Selection Effects} \label{sec:selection}

In this section, we investigate some possible biases in our
sample, and test whether these selection effects 
affect the conclusions of this paper.
 
\subsection{The Limited Field of View} \label{sec:FOV}

\begin{figure}[ht!]
\plottwo{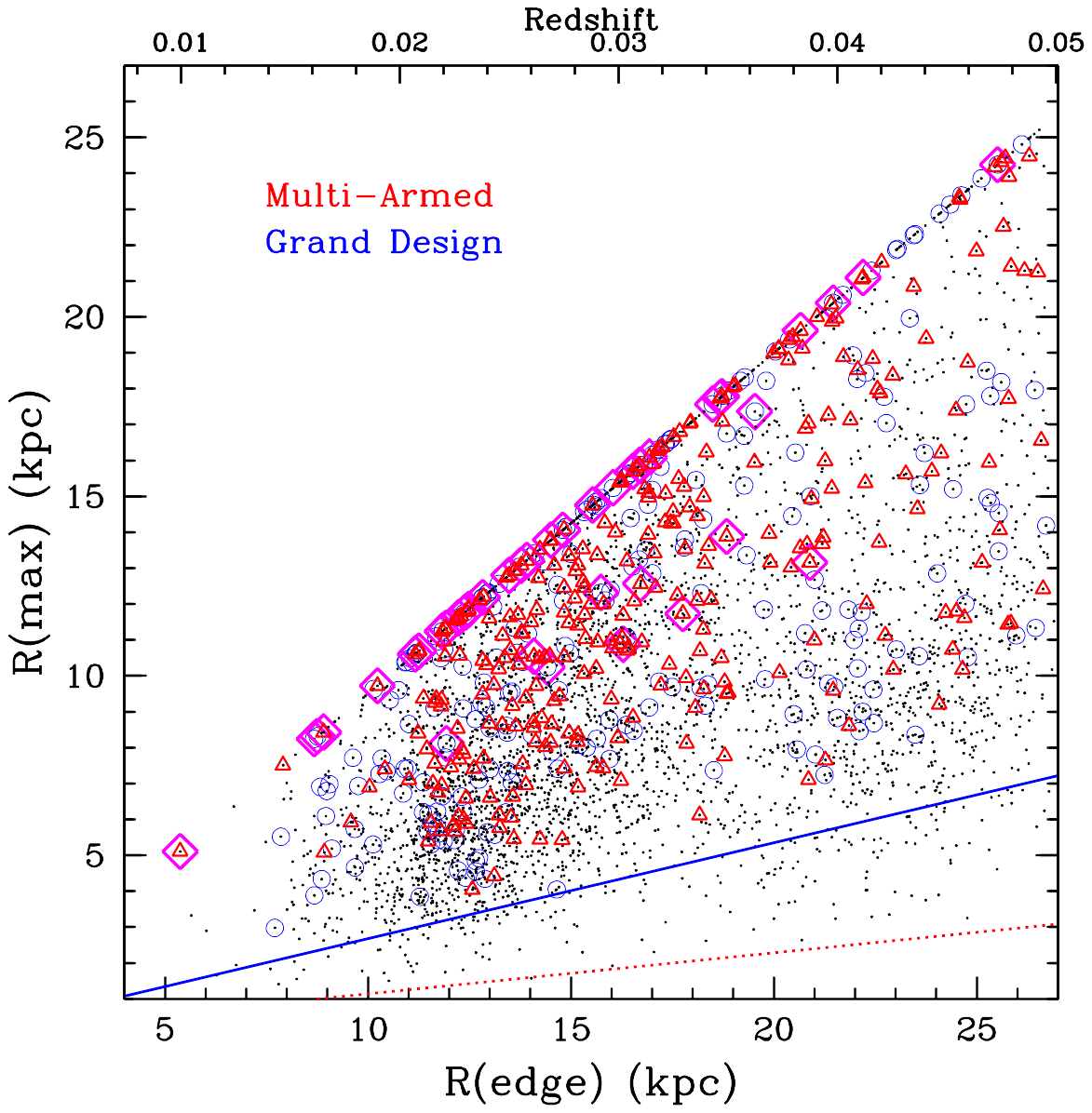}{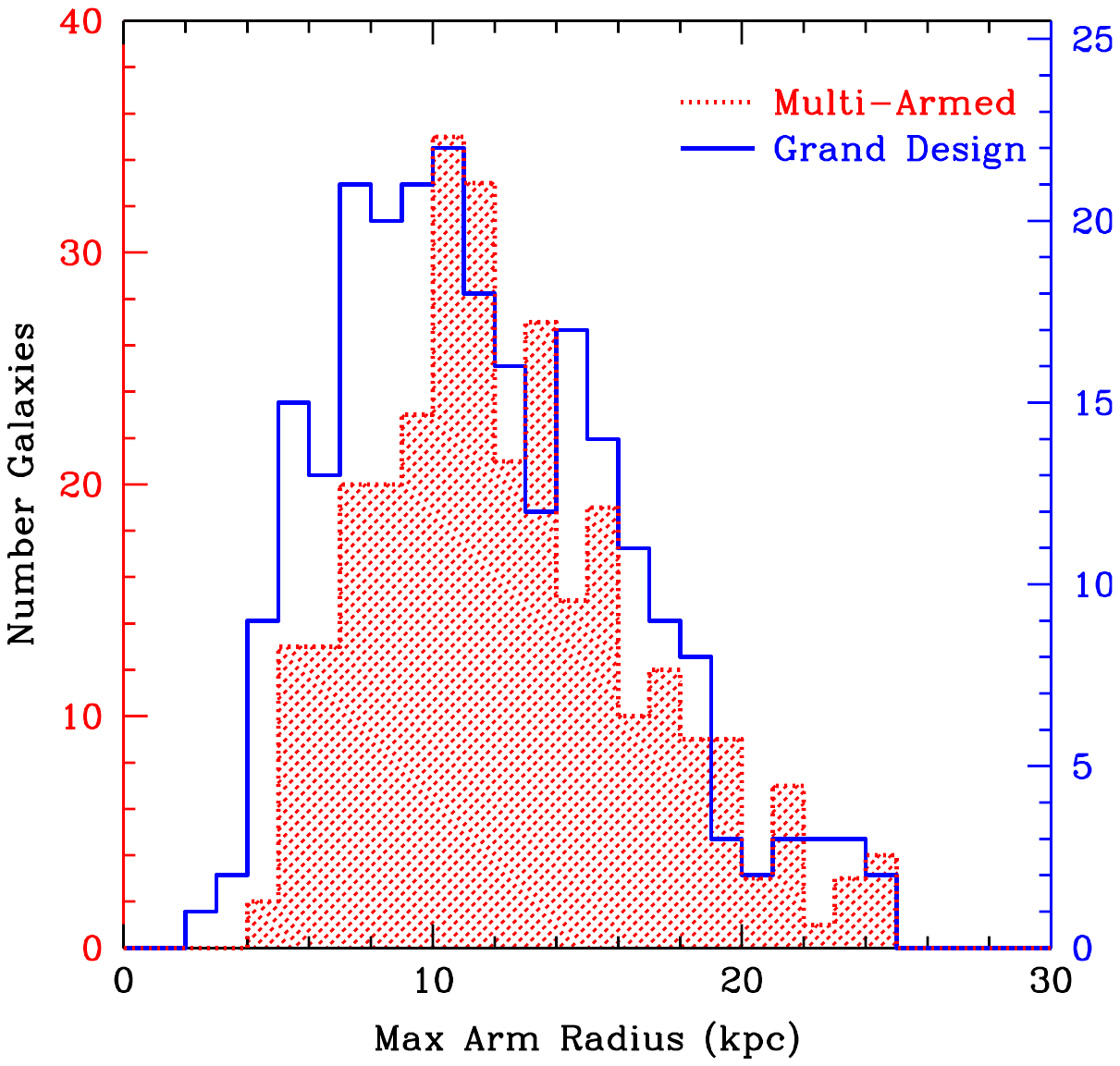}
\caption{
Left:
Plot of R(max), the maximum distance out to which we trace
spiral arms, vs.\ R(edge), the edge of the GZ-3D field of view.
The small black dots are 
all of the galaxies in the z $<$ 0.05 GZ-3D sample for which we are able
to mark spiral arms.
Red open triangles mark the galaxies that we classified as
multi-armed.  
Blue open circles are galaxies classified
as grand design. 
Magenta open diamonds
mark galaxies which have R90/R(edge) $>$ 1.2.
These are galaxies 
that are physically large compared to the GZ-3D field of view.
The red dotted line is the minimum R(max) possible for
galaxies flagged as 3-armed.
The blue solid line is the minimum R(max) possible for a galaxy defined
as grand design by our definition.
Right: Histograms of the maximum radius out to which we are able to 
reliably mark spiral arms (i.e., at a threshold = 3)
for
the multi-armed galaxies (red hatched histogram)
and the grand design galaxies (blue).
\label{fig:RmaxRedge}}
\end{figure}

Our arm-tracing method is limited by the 52$''$ $\times$ 52$''$
GZ-3D field of view.  This means that we are not able to measure arms
out very far in nearby galaxies.  
In the left panel of Figure \ref{fig:RmaxRedge}, we provide a plot
of the maximum radial distance to which we measure arms, 
R(max), vs.\ R(edge), the
radius to the edge of the observed field of view,
for all of the galaxies in the low z GZ-3D sample.
R(max) is defined as the maximum radial extent of an arm
above our threshold = 3 cutoff in the GZ-3D masks.
R(edge) in kpc is proportional to redshift, which is marked 
on the top axes of these plots.
The distributions of galaxies we identify as
multi-armed (red open triangles)
is similar to that of the 
grand design galaxies (blue open squares).
In this plot, we also identify galaxies 
with NSA R90 sizes 
that are 
greater than 1.2 $\times$ R(edge). 
These are 
galaxies with large physical radii compared to the field of
view, and so are more likely to have arms that are
truncated by the field of view.
The series of datapoints marking the
one-to-one relation which bisects these plots marks the maximum
distance we are able to measure arms.   
We are able to trace arms to greater extents 
in more distant galaxies, and therefore our arm measurements
may be artificially
truncated for nearby and/or physically large galaxies. 
This potentially causes a redshift bias in our study, in that our method
of counting arms and flagging
galaxies searches an increasingly large fraction of the galaxian disk
at increasing redshift.

In the right panel of Figure \ref{fig:RmaxRedge},
we provide histograms of R(max), the maximum radius in kpc
out to which
we identify spiral arms. 
There is a large spread in R(max) for the galaxies
in the sample, from 2 kpc to 25 kpc.
Although
the peak in the histogram of R(max) 
is slightly lower for grand design galaxies 
than for multi-armed galaxies,
there
is only a marginal statistical difference between the two 
samples (KS probability = 0.038).
Dividing the sample up into narrow bins of M* (0.3 dex bins),
we see an increase in R(max) with M*, but 
within a given M* bin KS tests do not show a significant difference
between grand design and multi-armed galaxies.

\begin{figure}[ht!]
\plottwo{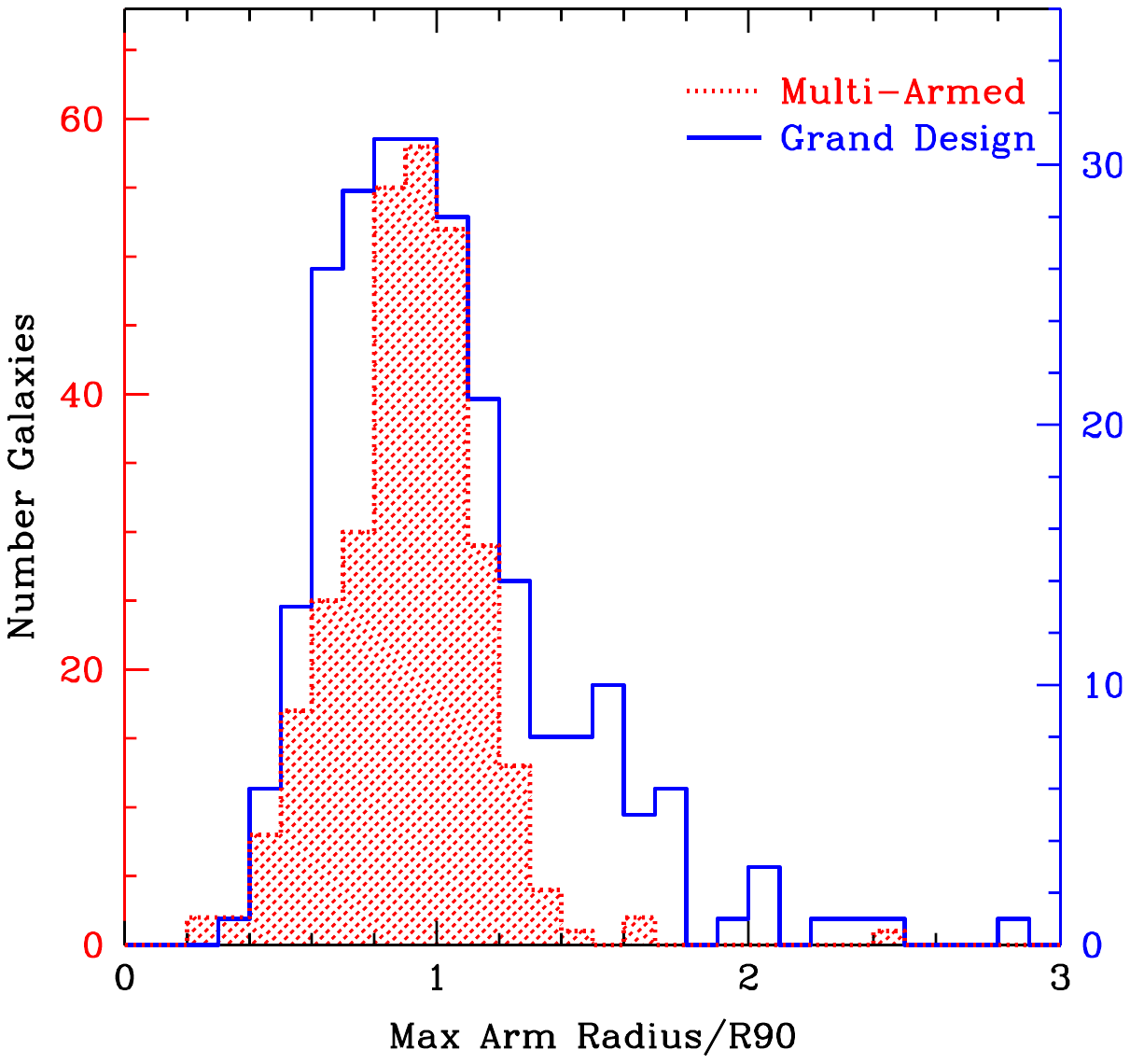}{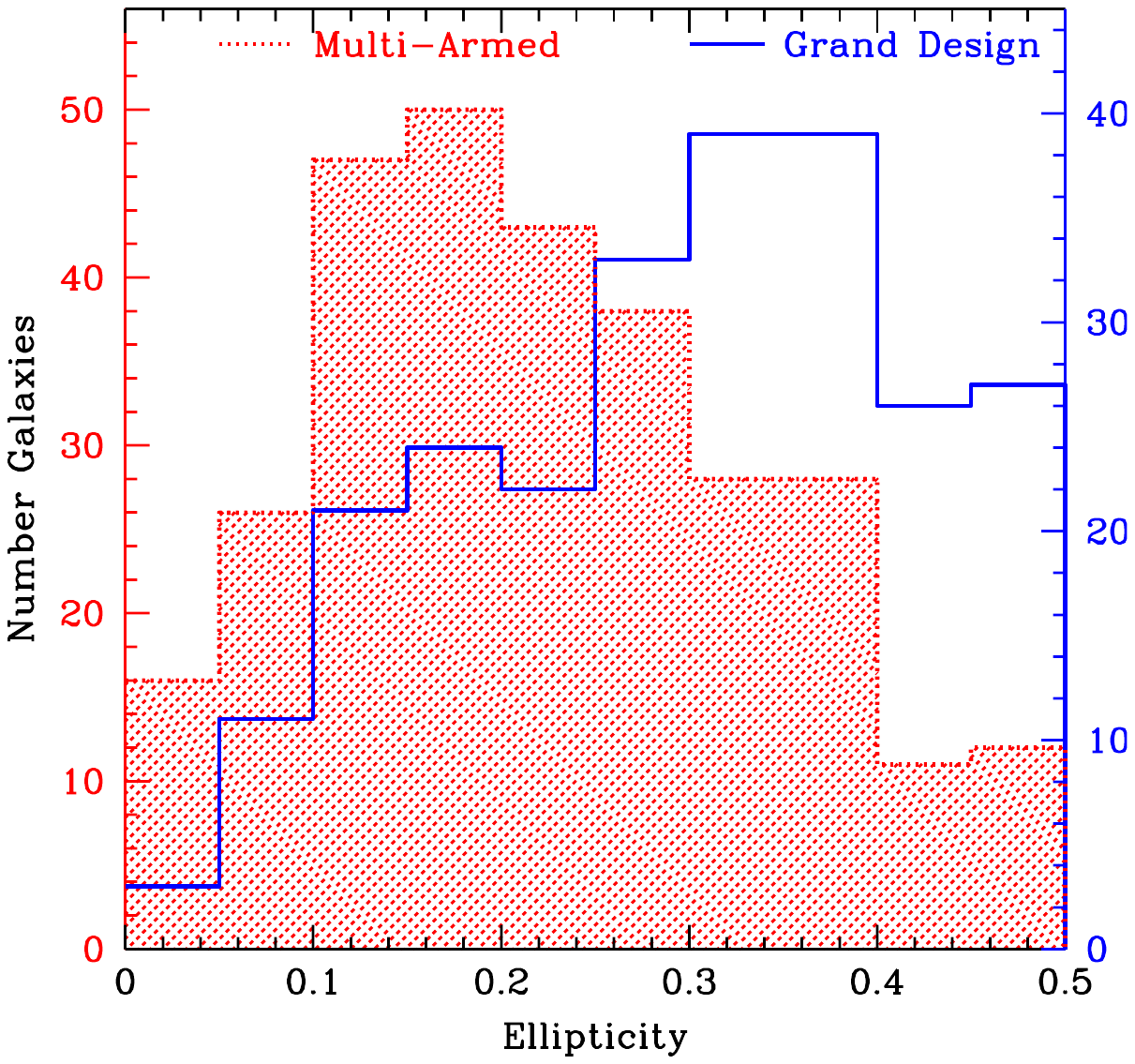}
\caption{ 
left: 
Histograms of R(max)
divided by R90, where R90 is the 
radius enclosing 90\% of the light.
Right: histograms of the NSA ellipticities for
the multi-armed galaxies
(red hatched) and grand design galaxies (blue).
\label{fig:maxarm_ellipticity}}
\end{figure}

In the left panel of 
Figure \ref{fig:maxarm_ellipticity}, we plot R(max) normalized by the R90 radius.
The peak of the distribution of R(max)/R90 
for the grand design galaxies is lower 
than for the multi-armed galaxies.
A KS test shows a significant difference for the full M* range
(KS probability 0.00048).
Sub-dividing into small mass bins, significant differences are present for
the 10 $\le$ log (M*/M$_{\sun}$)  $<$ 10.3 and 
10.3 $\le$ log (M*/M$_{\sun}$) $<$ 10.6 
bins
(KS probabilities of 0.00013 and 0.00375, respectively), with the
grand design galaxies having larger ratios.
The grand design galaxies in these bins
have median 
R(max)/R90 values larger than one (1.10 and 1.11, respectively).
The GZ-3D
participants were able to trace the spiral
arms out to large distances in the grand design galaxies.
Why there is this difference between grand design and multi-armed galaxies
is uncertain.  
One possibility is that the arms are stronger and more distinct
in grand design galaxies, making them easier to see further out in the 
galaxies.  However, we don't see a significant difference in
arm strengths between grand design and multi-armed galaxies
(Figure \ref{fig:f3}).

To investigate possible selection effects due to spiral arms extending
beyond the edge of the field of view, we re-ran the KS tests 
comparing stellar mass, concentration, and sSFR 
for the two samples
of galaxies,
after
eliminating galaxies with 
R(max) $\ge$ 0.95 R(edge) (i.e., galaxies that may have truncated
arms in GZ-3D).
The differences in stellar mass and concentration
between grand design and multi-armed galaxies persist,
indicating that these differences are not caused
by selection effects from
the limited field of view.  
We do not see significant differences in the sSFRs of the grand design
and multi-armed galaxies when binning by C and M* as in 
Figure \ref{fig:mainseq},
after
eliminating galaxies with R(max) $\ge$ 0.95 R(edge). 
Thus our basic conclusions about M*, C, and sSFR do not change
when eliminating galaxies that may have arms that are truncated
by the limited field of view.

\subsection{Inclination} \label{sec:inclination}

We also checked to see if the inclination of the galaxy
(i.e., the ellipticity of the SDSS image) biases our classification into
grand design vs.\ multi-armed.
In the right panel of
Figure \ref{fig:maxarm_ellipticity}, we provide histograms of the
NSA ellipticity of 
our multi-armed and grand design galaxies.
There is a clear difference, with the galaxies classified
as multi-armed having smaller ellipticities than
the grand design galaxy.
In other words, the
multi-armed galaxies in our sample are more face-on on average than the
grand design galaxies.
This result is confirmed by a KS test (KS probability
10$^{-10}$).

The reason for the observed difference between the
two samples 
is uncertain.
It may be harder to see all of the arms if the galaxy is inclined,
which may cause viewers to sometimes miss arms,
leading to an undercount of the arms in multi-armed galaxies.
Viewers may also be more likely to not to mark any arms at all
in inclined galaxies,
causing the galaxy not to be included in our final sample.
Another possibility
is that the ellipticity of a galaxy as determined 
from the outer optical isophotes may be over-estimated for
galaxies with two dominant strong arms.

To test whether this difference in inclination between the  
two samples is responsible for the observed differences in M* and C
between the two samples and the lack of an observed difference in
sSFR, we re-ran the KS tests 
for M*, C, and sSFR, 
limiting the sample to only the most face-on galaxies
(ellipticity $<$ 0.25).
The main results 
still
hold with these face-on sub-samples.
The only difference is that,
when the ellipticity is restricted to $<$0.25,
no significant differences in concentration 
between grand design
and multi-armed galaxies are seen
for the 9.4 $\le$ log (M*/M$_{\sun}$) $<$ 9.7 
bin.  A difference
in concentration
is still present in the higher mass bins.
No differences in sSFR are seen between the two groups
after the galaxies are binned in both C and M*, even after 
limiting the sample to galaxies with 
ellipticity $<$ 0.25.

\subsection{Incompleteness} \label{sec:incompleteness}

\begin{figure}[ht!]
\plottwo{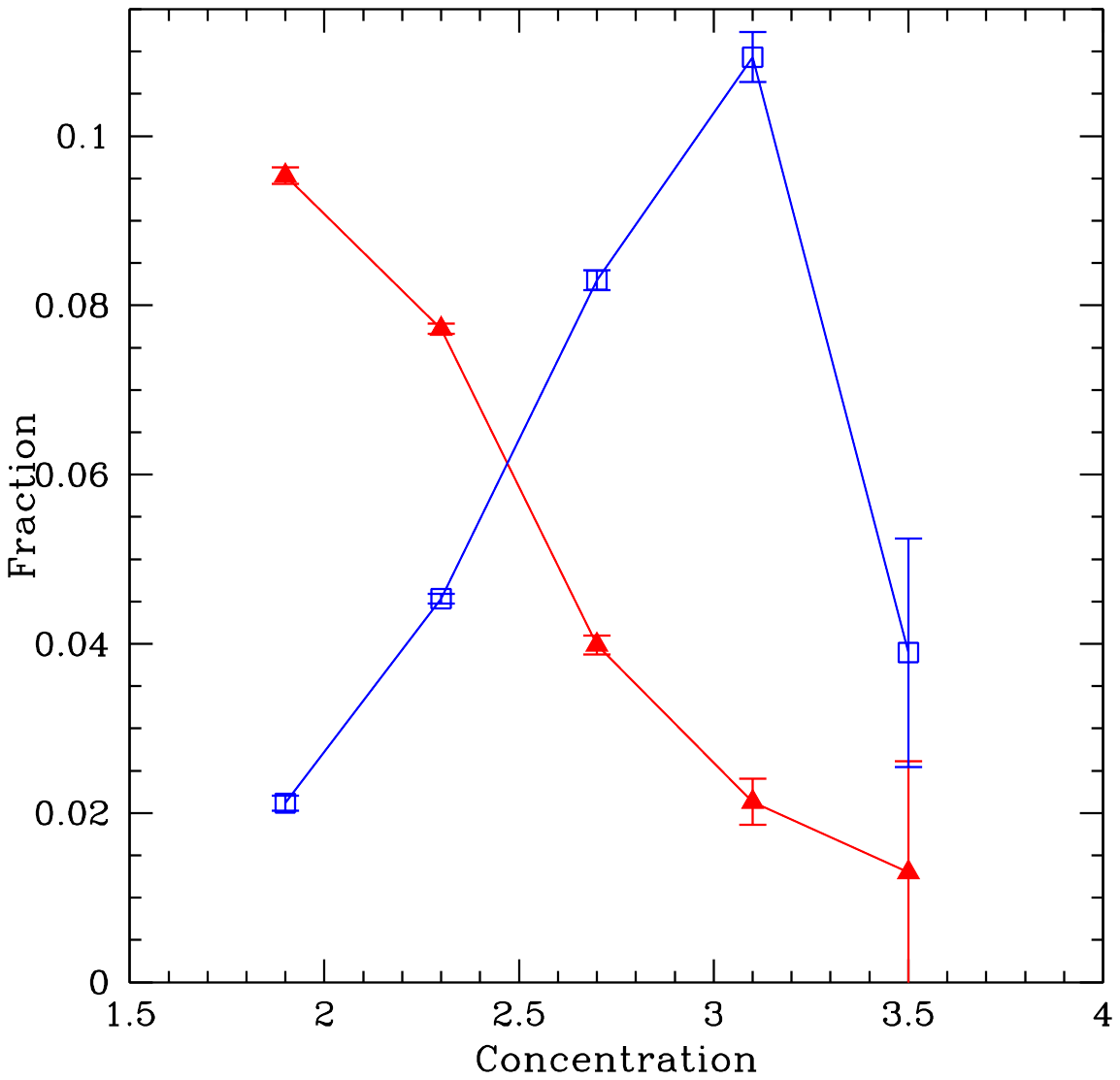}{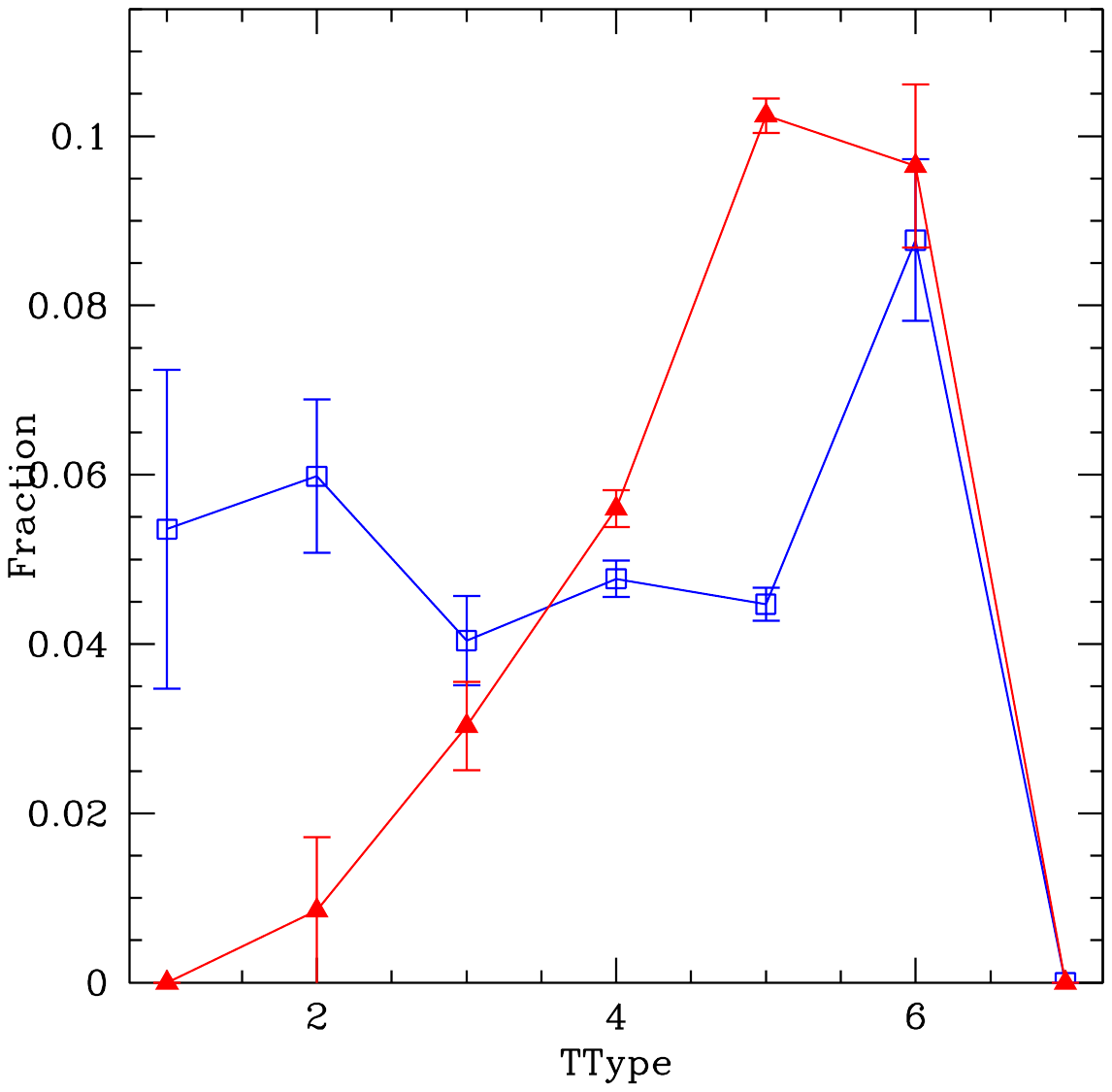}
\caption{ 
The fraction of the z $<$ 0.05 GZ-3D galaxies
in our final samples of multi-armed galaxies (red filled triangles)
and grand design galaxies (blue open squares),
as a function of concentration (left panel) and 
TType (right panel).
The uncertainties are $\sqrt{N}$.
\label{fig:fractionC}}
\end{figure}

Our method of identifying multi-armed and grand design galaxies
is quite incomplete; only $\sim$12\% of the original
sample of 4449 z $<$ 0.05, ellipticity $\le$ 0.5 GZ-3D spiral arm galaxies
end up in either of the final classes.
In the left panel of Figure \ref{fig:fractionC}, 
we plot the fraction of galaxies in our original GZ-3D sample that
were 
identified in this study 
as multi-armed and as grand design,
as a function of concentration.
At low concentrations,
the multi-armed galaxies are a larger fraction 
of the total z $<$ 0.05 population of GZ-3D galaxies.  
In the middle of the concentration range,
the fractions that are grand design galaxies
are higher.
At all concentrations, 
however, only a small fraction of galaxies
ended up in either of our final
sample.   The incompleteness is particularly high at large C.

For the galaxies in our original low z GZ-3D sample
that have TTypes from 
\citet{2018MNRAS.476.3661D},
in the right panel of Figure \ref{fig:fractionC} we plot
the fraction of galaxies identified in this study as
multi-armed or grand design, as a function of TType.
There is a clear difference between the multi-armed
and grand design galaxies in this plot, with the grand design
galaxies skewed to earlier TTypes.
However,
only a small fraction of the galaxies classified
as 1 $\le$ TType $<$ 7 ended up in either
our grand design or multi-armed sample.
The incompleteness is highest at TType = 7 and TType = 1.

\begin{figure}[ht!]
\plottwo{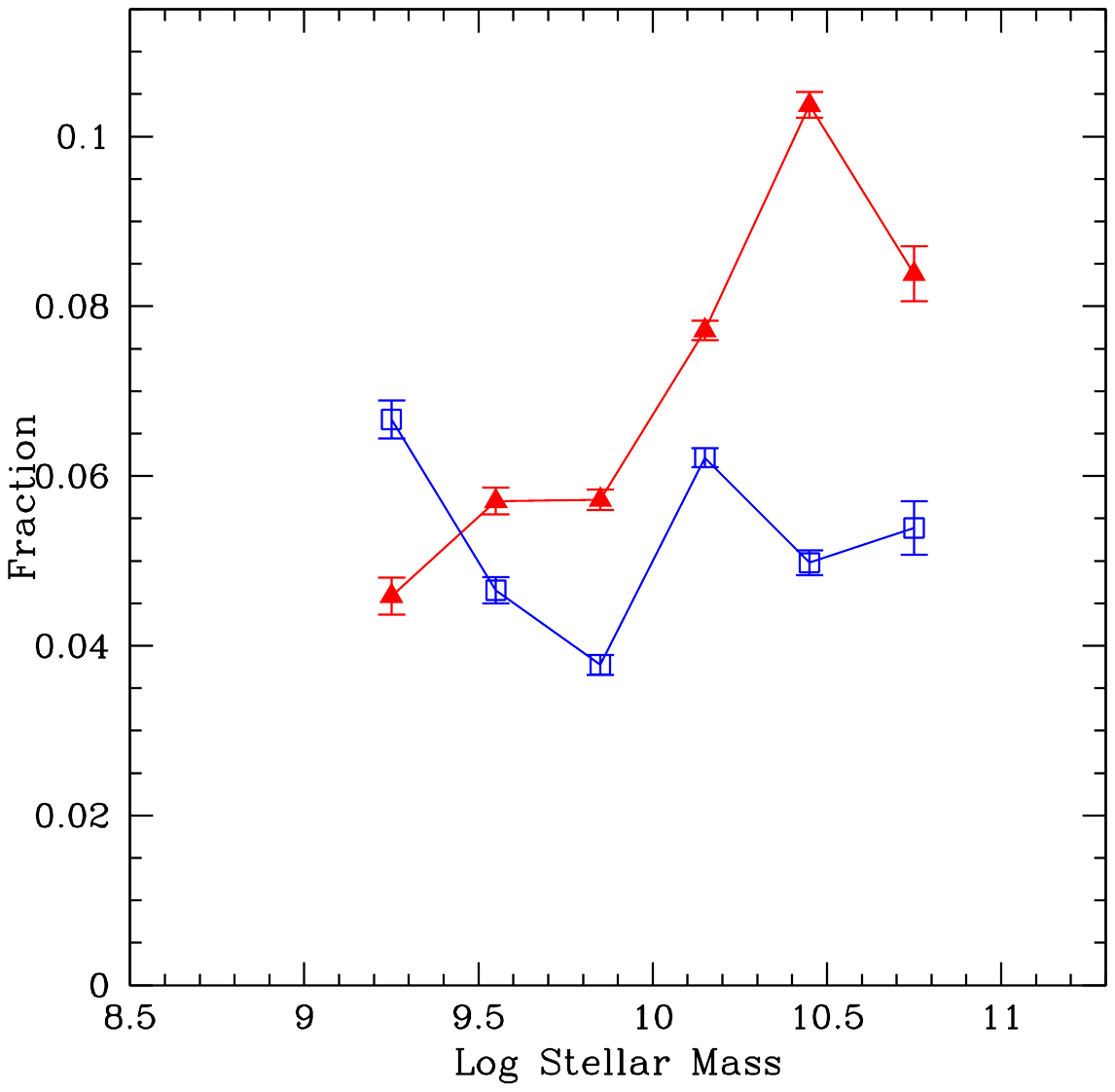}{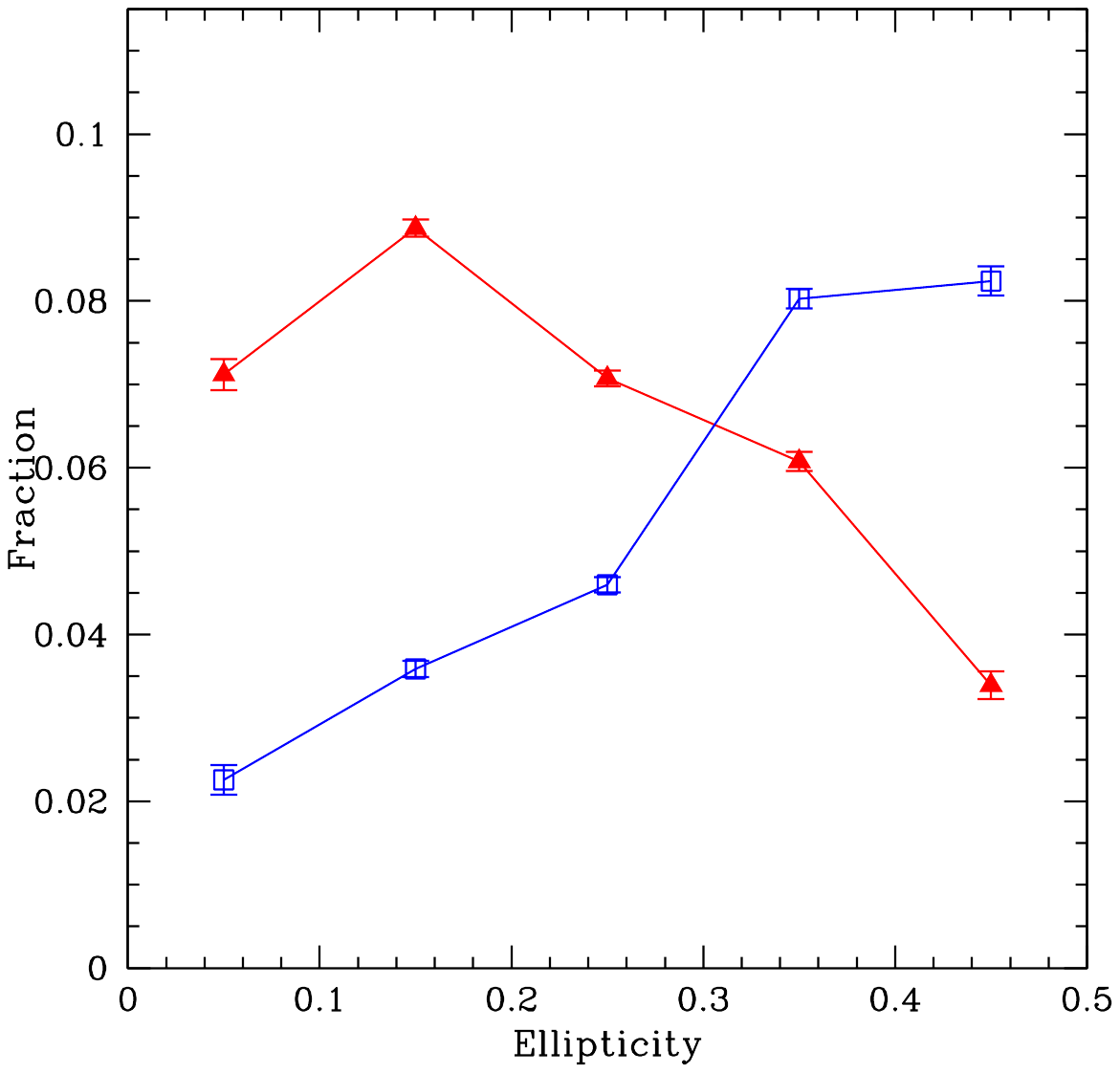}
\caption{ 
The fraction of the z $<$ 0.05 GZ-3D galaxies
in our final samples of multi-armed galaxies (red filled triangles)
and grand design galaxies (blue open squares),
as a function of stellar mass (left) and ellipticity (right).
The uncertainties are $\sqrt{N}$.
\label{fig:fractionMstar}}
\end{figure}

In the left panel of Figure \ref{fig:fractionMstar}, we plot the fraction
of the original low z GZ-3D sample as a function of M*.
Except for the lowest mass bin, 9.1 $\le$ log (M*/M$_{\sun}$) $<$ 9.4, 
there are more multi-armed galaxies than grand design.
The ratio of multi-armed to grand design is highest at
10.3 $\le$ log (M*/M$_{\sun}$) $<$ 10.6. 
The sample is more complete at higher masses.
The right panel of Figure \ref{fig:fractionMstar}
displays completeness vs.\ ellipticity for the two classes
of spirals.  As noted earlier, there is a bias in that our
multi-armed galaxies appear more face-on than the grand design
galaxies.  

Several factors may contribute
to the observed incompleteness.
First, insufficient resolution and/or sensitivity
in the SDSS images of some galaxies
may have caused the GZ-3D viewers to not mark
the spiral arms, or to mark the arms incorrectly.
Second, although the MaNGA GZ-3D sample was selected to avoid
flocculent galaxies, 
some galaxies may have too short
or too fragmented arms to be marked by the GZ-3D participants,
or the spiral segments
marked by the GZ-3D participants
may be too short to meet our criteria for flagging.
Third, in some cases the arms may have been hard to pick out
in the images,
having high pitch angles and/or low arm/interarm contrast. 
Fourth, 
some galaxies may not fit cleanly into
either class. For example, some galaxies may be too peculiar
and/or strongly interacting.
Fifth, some GZ-3D viewers may have only marked two arms per
galaxy, and may not have marked branches.
This may cause multi-armed galaxies to end up in our grand design
sample, and then be eliminated by the by-eye inspection.

To better understand the evolutionary relationship between
grand design and multi-armed galaxies, it would be useful
to determine the arm structures of all of the galaxies
in the full z $<$ 0.05 GZ-3D sample.
Obtaining clear morphological information for a more complete
sample of spirals will help us better understand the cause and evolution
of spiral patterns in galaxies.
In the future,
it would be worthwhile repeating the current study with
higher resolution, more sensitive images 
from more modern surveys
such as the Dark Energy Survey or the Euclid telescope.

\bibliography{arm_count_paper}{}
\bibliographystyle{aasjournal}



\end{document}